\documentclass[aps,twocolumn,superscriptaddress,nobibnotes,showkeys]{revtex4-1}

\usepackage{times}
\usepackage[utf8]{inputenc}
\usepackage{amsmath}
\usepackage{amssymb}
\usepackage{graphicx}
\usepackage{bm}
\usepackage{subfigure}
\usepackage{dcolumn}
\usepackage{color}
\usepackage[normalem]{ulem} 

\newcommand{\betashear}{\beta_{\rm sh}}
\newcommand{\order}[1]{\mathcal{O}\left(#1\right)}

\newcommand{\volplasmafreq}{\omega_{\rm p}}
\newcommand{\surfplasmafreq}{\omega_{\rm sp}}
\newcommand{\tildesurfplasmafreq}{\tilde{\omega}_{\rm sp}}
\newcommand{\vfermi}{v_{\rm F}}
\newcommand{\halbeta}{\beta_{\rm H}(\omega)}

\renewcommand{\vec}[1]{\mathbf{#1}}
\renewcommand{\Re}{\operatorname{Re}}
\renewcommand{\Im}{\operatorname{Im}}

\newcolumntype{d}[1]{D{.}{.}{#1}}

\begin{document}

\title{Halevi's extension of the Euler-Drude model for plasmonic systems}

\author{Gino Wegner}
\email[E-mail address: ]{wegner@physik.hu-berlin.de}
\affiliation{Humboldt-Universität zu Berlin, Institut für Physik, 
             AG Theoretische Optik \& Photonik, 12489 Berlin, Germany}
\affiliation{Friedrich-Schiller-University Jena, Institute of Condensed Matter Theory and Optics  
             Max-Wien-Platz 1, 07743 Jena, Germany}

\author{Dan-Nha Huynh}
\affiliation{Humboldt-Universität zu Berlin, Institut für Physik, 
             AG Theoretische Optik \& Photonik, 12489 Berlin, Germany}

\author{N. Asger Mortensen}
\affiliation{Center for Nano Optics, University of Southern Denmark, 
             Campusvej 55, DK-5230~Odense~M, Denmark}
\affiliation{Danish Institute for Advanced Study, University of Southern Denmark, 
             Campusvej 55, DK-5230~Odense~M, Denmark}

\author{Francesco Intravaia}
\affiliation{Humboldt-Universität zu Berlin, Institut für Physik, 
             AG Theoretische Optik \& Photonik, 12489 Berlin, Germany}
						
\author{Kurt Busch}
\affiliation{Humboldt-Universität zu Berlin, Institut für Physik, 
             AG Theoretische Optik \& Photonik, 12489 Berlin, Germany}	
\affiliation{Max-Born-Institut, 
             12489 Berlin, Germany}											
											
\date{\today}

\begin{abstract}
The nonlocal response of plasmonic materials and nanostructures is usually described within a 
hydrodynamic approach which is based on the Euler-Drude equation. In this work, we reconsider
this approach within linear response theory and employ Halevi's extension to this standard 
hydrodynamic model.
After discussing the impact of this improved model, which we term the Halevi model, on the 
propagation of longitudinal volume modes, we accordingly extend the Mie-Ruppin theory. 
Specifically, we derive the dispersion relation of cylindrical surface plasmons. 
This reveals a nonlocal, collisional damping term which is related to earlier phenomenological 
considerations of limited-mean-free-path effects and influences both, peak width and amplitude
of corresponding resonances in the extinction spectrum. 
In addition, we transfer the Halevi model into the time-domain thereby revealing a novel, diffusive 
contribution to the current which shares certain similarities with Cattaneo-type currents and analyze 
the resulting hybrid, diffusive-wave-like motion. 
Further, we discuss the relation of the Halevi model to other approaches commonly used in the literature.
Finally, we demonstrate how to implement the Halevi model into the Discontinuous-Galerkin Time-Domain 
finite-element Maxwell solver and are able to identify an oscillatory contribution to the diffusive 
current. 
The Halevi model thus captures a number of relevant features beyond the standard hydrodynamic
model. Contrary to other extensions of the standard hydrodynamic model, its use in time-domain 
Maxwell solvers is straightforward -- especially due its affinity to a class of descriptions that 
allow for a clear distinction between bulk and surface response. This is of particular importance 
for applications in nano-plasmonics where nano-gap structures and other nano-scale features have 
to be modeled efficiently and accurately.
\end{abstract}


\maketitle

\section{\label{sec:introduction}Introduction}

Over the last decades, the preparation of complex metallic nanostructures has experienced 
tremendous progress due to advances in material quality and nanostructuring techniques.  
Examples are dimers of spheres
~\cite{https://doi.org/10.1002/qute.201800016}
and cylindrical wires
~\cite{PhysRevB.95.115441},
mixed wire-sphere systems
~\cite{doi:10.1021/acsnano.0c05240}, and bowtie antennas. 
The latter are connected 
~\cite{doi:10.1021/nn402323t} 
or unconnected
~\cite{PhysRevB.72.165409}, 
tip-to-tip nanotriangles. 
All such dimers (and many more conceivable structures) feature nano-gaps
that give rise to 
strong field enhancements that can be exploited for numerous applications which include 
(but are certainly not limited to) sensing via surface- and tip-enhanced Raman scattering 
and surface-enhanced infrared absorption spectroscopy as well as various wave mixing processes.   
The performance of such devices relies on the plasmonic properties of the conduction electrons 
at optical frequencies and they can be taylored via the choice of material as well as via
size and geometry
~\cite{doi:10.1021/nl073042v}
of the system's constituents as well as on their surface preparation
~\cite{PhysRevB.9.1277,wegner2020remarks}.

The efficient modeling of plasmonic nanostructures on both, the interpretative and predictive
level requires appropriate material models that are able to capture the relevant physics on
the involved time- and length-scales and are amenable to a performant implementation in
numerical scheme. Owing to the dispersive nature of the plasmonic material response, the 
latter aspect is particularly important for time-domain simulations of the Maxwell equations.
One particular semiclassical model, which satisfies these requirements is often referred to
as the hydrodynamic 
~\cite{electron_density_hydro_profile,PhysRevB.85.201403,PhysRevB.86.115451,dhuynh2016} 
or hydrodynamic Drude model 
~\cite{HIREMATH20125890, PhysRevB.97.075431,Mortensen_meso_edyn_at_surf_2021}
whose details we describe in section 
\ref{sec:justification_halevi_model}. 
On the one hand, it captures certain quantum effects which become relevant in the nanometer 
regime and, on the other hand, its coarseness allows for a language of just a few degrees 
of freedom. 
Therefore, when being incorporated into a Maxwell solver, it facilitates the treatment of
scattering phenomena of nano-scale features that modulate a much larger structure. In 
practise, this multi-scale aspect is accompanied by a moderate usage of computational 
resources.
  
An illustrative phenomenon which is well-described by the original hydrodynamic model is 
the size-dependent shift of plasmonic resonances for small metallic particles 
~\cite{RUPPIN2001205,resonance_shifts_spill_out_hydro, first_GNOR_paper,Raza_2015}. 
Further, through the addition of diffusive dynamics the size-dependent broadening of 
these resonances is qualitatively accounted for within the so-called Generalized Nonlocal 
Optical Response (GNOR) model
~\cite{first_GNOR_paper,Raza_2015}. 
In the GNOR model, the standard hydrodynamic current is supplemented by a diffusive 
current that follows Fick's first law. Furthermore, comparing the GNOR model to Halevi's 
extension of the hydrodynamic model (which is based on the Boltzmann-Mermin approach)
facilitates the derivation of a quantum mechanical diffusion constant.

In this work, we will reconsider the plasmonic response of metallic structures by direct 
application of Halevi's extension of the hydrodynamic model which we refer to as the 
Halevi model. 
As we go along, we also focus on various extensions to the original hydrodynamic model and 
elaborate on the difficulties of a one-to-one identification with the GNOR model. While 
the Halevi as well as the GNOR model initially describe the dynamics of electrons in the
bulk, we determine further reponse  properties near the surface of a finite scatterer.
Specifically, we consider nano-scale wires as prototypical structures. 
In fact, nano-wires are rather popular and convenient for experimental studies as they
can be fabricated with excellent quality via membrane- or template-based synthesis with
diameters down to about $5$nm
~\cite{doi:10.1126/science.266.5193.1961}. 
Possible nano-scale devices, that utilize plasmonic nano-wires include waveguides
~\cite{PhysRevLett.93.137404,doi:10.1021/jp001435b},
nano-antennas
~\cite{doi:10.1021/nl073042v},
and sensors
~\cite{doi:10.1021/acsnano.0c05240}.
. 

Under normal illumination, analytical expressions for the electromagnetic fields in 
and around infinitely-extended straight plasmonic nano-wires described by the hydrodynamic 
Drude model have been obtained with
~\cite{VillPrez2009HydrodynamicalMF} 
and without
~\cite{RUPPIN2001205}
the quasi-static approximation.
Here, the incident plane wave's electric field 
is polarized perpendicular to the nano-wire axis
~\cite{VillPrez2009HydrodynamicalMF,RUPPIN2001205}. 
In particlar, we aim at extending the work of Ref.~\onlinecite{RUPPIN2001205} 
to the Halevi model. 

The manuscript is organized as follows. 
Starting with Sec.~\ref{sec:justification_halevi_model} we briefly review the basics 
of the hydrodynamic Drude model and elaborate on the extension derived by Halevi. 
Further, we place this description into the framework of the viscoelastic model
~\cite{PhysRevB.60.7966} 
in order to deepen the conceptual understanding and provide a route for future extensions. 
In the following Sec.~\ref{sec:Ruppin_cyl_halevi}, we adapt the theory developed in Ref.~\onlinecite{RUPPIN2001205} 
to the Halevi model, provide a corresponding justification, and elaborate on
the implications. In particular, we discuss the Halevi model's influence on mode 
propagation and derive the dispersion relation of surface plasmon polaritons within 
the wire. The latter reveals a nonlocal damping term that is connected to an earlier
phenomenological proposition on limited-mean-free-path effects and resonance 
broadening. In the subsequent Sec.~\ref{sec:deriv_drift_diffusion_current} we develop
the time-domain formulation of the Halevi model. We explicitly show, that the Halevi
extension to the hydrodynamic model introduces a novel current, that is conceptually
closely related to the Cattaneo-current
~\cite{Compte_1997} which models classical diffusion processes 
with finite propagation velocity. 
Next, in Sec.~\ref{sec:comparison_Halevi_GNOR} we briefly review the GNOR-model to 
allow for a detailed comparison with the Halevi model, specifically regarding the 
respective diffusive paradigms. We will show that, despite certain similarities, 
there are marked differences between the two models which prohibit a one-to-one 
mapping. In Sec.~\ref{sec:halevi_phase_shift_simulations} we then proceed to 
numerical simulations of the scattering setup of Sec.\,\ref{sec:Ruppin_cyl_halevi} 
using the time-domain formulation of the Halevi model and analyze the influence 
of the Halevi extension on the propagation of the electric field and the mode 
profiles. Finally, in Sec.~\ref{sec:summary_and_outlook} we summarize our findings
and provide an outlook for future studies.
 
\section{\label{sec:justification_halevi_model}Justification of the Halevi model}

It appears, that Felix Bloch proposed and discussed the first treatment of electron 
dynamics by means of a hydrodynamic model
~\cite{PhysRevB.51.7497} 
(see also Refs.~\onlinecite{Raza_2015,Ying_hydro_response_inhom_metal,PhysRevB.60.15550,Aers1980NonradiativeSP}
for further discussions)
Back in 1933, as demonstrated in Ref.~\onlinecite{bloch_bremsvermoegen}, such an approach 
presented an analytically amenable means to estimate the stopping power associated
with the response of certain atoms.
Since then, the model has been rederived and extended following different paradigms. 
Usually the connection to conservation equations is pointed out and their specific
forms are related to an equation of state. The latter may be deduced from (quantum) 
statistics
~\cite{PhysRevB.60.15550,haas2011quantum,ancona_hd_models_semicond,doi:10.1137/S0036139992240425}. 
A special case is the derivation of equations of motion from an energy principle. 
Within such an approach, the equation of state follows from the choice of internal 
energy functional 
~\cite{bloch_bremsvermoegen,doi:10.1063/1.5003910,resonance_shifts_spill_out_hydro,Ying_hydro_response_inhom_metal, PhysRevB.91.115416}. 
Another strategy considers already existing, semi-classical models and tries to 
asymptotically identify their response functions with those which are motivated by 
continuum theories 
~\cite{PhysRevB.51.7497, PhysRevB.60.7966,PhysRevB.91.115416, universe7040108}.
In the present work, we follow the latter approach. 

The continuum assumption lies at the heart of hydrodynamic models for describing the
conduction electrons in metals. This means that a mesoscopic perspective is adapted 
where the electrons form a charged fluid such that a given fluid element is (i) much 
larger than the actual constituents of the fluid and their mean separation and (ii) 
much smaller than the volume occupied by the fluid
~\cite{fluid_mechanics}. 
The ionic background is treated as a rigid, motionless continuum -- the rigid jellium
~\cite{giuliani_vignale_2005}-- and restores overall charge neutrality. The thus introduced
electron continuum may dynamically change shape and volume. 

The associated dynamical 
quantities are the moments of a distribution function that describes the microscopic 
electron dynamics. 
Quite generally, this distribution function $f(\vec{r}, \vec{p}, t)$ represents the
probability of finding at time $t$ a representative electron with its microscopic 
momentum $\vec{p}$ in an infinitesimal volume element surrounding the position 
$\vec{r}$. 
{
This effective one-particle distribution function is obtained from the distribution 
function of the full $N$-body-electron-system by integrating over the positions and 
microscopic momenta of the remaining $N-1$ particles. Due to the Coulomb interaction 
between the electrons, as well as their interaction with the ionic background, the 
dynamics of a $m$-particle-distribution contains terms that couple to the $m+1$-particle 
distribution where $0<m<N$. This coupled set of $N-1$ equations is referred to as the 
Bogoliubov–Born–Green–Kirkwood–Yvon hierarchy
~\cite{haas2011quantum,PhysRevLett.128.190401}. Breaking this hierarchy then facilitates 
the implementation of the above-mentioned method of moments.
Among the truncated descriptions of one-particle distributions, a suitable equation 
of motion, from which the hydrodynamic equations can be deduced is the  
Boltzmann equation 
~\cite{Kittel1963,ashcroft1976solid} 
with an electron-ion collision integral that may be expressed following Mermin's recipe for a 
charge-conserving, single-relaxation-time correction
~\cite{PhysRevB.1.2362} 
and the electron Coulomb interaction being treated by a mean-field approximation.
Upon expanding the distribution function of the Boltzmann-Mermin model into the moments
of the microscopic momentum another hierarchy of evolution equations is obtained where 
successive-order moments become coupled and the individual equations of motion represent 
conservation laws.
A closed set of equations of motion can be obtained through further approximations that 
relate a given moment to lower-order moments only -- specifically, this procedure is 
connected to additional approximations to the chosen coupling term of the distributions 
functions. 

The standard hydrodynamic
model considers the first- and second-order moments, charge- and momentum density,
and the corresponding equations of motion ensure charge conservation and momentum 
balance. 
In the resulting evolution equations, the momentum density is
often replaced by the charge current density. The omission of higher-order moments 
and the corresponding equations of motions is tied to the absence of heat conduction
~\cite{ancona_hd_models_semicond} 
and to a specific form of the stress tensor (a.k.a. momentum current density). The 
latter is given by a pressure that can solely be expressed in terms of the charge 
density which, e.g., may be derived from the Thomas-Fermi theory of a degenerate 
electron gas
~\cite{bloch_bremsvermoegen}, thereby terminating the aforementioned second hierarchy 
and arriving at a closed set of equations. 
This description provides a starting 
point in the sense that it incorporates quantum-statistical effects in the kinetic 
energy of the electronic many-body system, eventually leading to a nonlocal reponse, 
while it excludes certain effects such as the exchange-correlation dynamics which, 
at least in three-dimensional systems with sufficiently high densities of free 
electrons, can be neglected to a first approximation.
}
 
Different choices for the stress tensor are possible depending on the desired level 
of accuracy and on the particular physical system. In principle, more sophisticated 
models could be introduced, leaving the coupling term of momentum and energy density 
untouched. Then, a constitutive equation for the heat conduction could be provided
as it would otherwise couple to even higher-order equations beyond the conservation 
of energy 
~\cite{ancona_hd_models_semicond}.

Specifically, the equation of motion of the charge density is dictated by the 
continuity equation
\begin{align}
   \partial_t \rho(\vec{r}, t) + \nabla \cdot \vec{J}(\vec{r}, t) & = 0.
\label{eq:continuity_eq_time_position}
\end{align}  
Due to absence of source and sink terms, the above equation stipulates 
that charges are neither produced nor annihilated. 

The conservation of the current density $\vec{J}$ (or, equivalently, momentum conservation) 
is given by the linearized Euler equation of classical fluid dynamics 
~\cite{fluid_mechanics}
and reads
\begin{align}
  \partial_t \vec{J}(\vec{r}, t) &= \epsilon_0\volplasmafreq^2 \vec{E}(\vec{r}, t) 
	                                  - \beta^2_{\rm TF} \nabla \rho(\vec{r}, t).
\label{eq:def_Euler_equation}
\end{align}
Here, we would like to note, that the strength of the Thomas-Fermi-pressure term scales 
with the parameter $\beta_{\rm TF}=\vfermi/\sqrt{3}$ and, thus with the Fermi velocity 
$v_{\rm F}$. It therefore inherits constraints of the Pauli exclusion principle 
~\cite{haas2011quantum}. 
In fact, the parameter $\beta_{\rm TF}$ may be viewed as the velocity of sound in the electron 
continuum 
~\cite{lindhard_prop_gas_charged_part} 
and, accordingly, characterizes density or pressure waves 
~\cite{first_GNOR_paper} 
which may build up in the electron continuum. The total current is defined via the 
center-of-mass motion which arises from the total electric field $ \vec{E} (\vec{r}, t)$ 
induced by the motion of the electron density relative to the fixed jellium background
and any externally applied electric fields. 
In this treatment, we neglect the effects of magnetic fields (see Ref.~\onlinecite{Wolff:13} 
for the treatment of magneto-optic effects). 
This classical treatment of the electric field excludes quantum-mechanical effects such 
as the exchange interactions. 
{
An early work, based on Bloch's hydrodynamic model, that derives resonances of a 
metallic scatterer without collisions but with an approximate form of electronic 
exchange-interactions, has been presented by Jensen in Ref.~\onlinecite{jensen_hd_and_exchange}.
}

Since Bloch's model focussed more on the electrons within the individual atoms/ions 
and less on the relative position of the latter, we expect deviations to his theory 
when considering more realistic metallic structures. As a first correction, the Drude 
term $-\gamma \vec{J}$ is added to the r.h.s. of Eq.~(\ref{eq:def_Euler_equation}) 
yielding the Euler-Drude model. 
This new term describes the hinderance of electronic motion due to aperiodicities in the 
ionic lattice that may come from defects in the crystal lattice or from lattice vibrations
(phonons). It leads to collisions quantified with a rate $\gamma$ which diminish electron 
momentum
~\cite{PhysRevB.60.7966} 
and is part of the first-order moment of the single-relaxation-time approximation, 
discussed above. 

As Bloch points out, the Thomas-Fermi model has been applied to the static behaviour of the 
electron continuum. From Eq.~(\ref{eq:def_Euler_equation}) we, thus, expect a quasi-static 
description. But such a treatment fails for typical metals at optical frequences. As a remedy, 
the very same equation may be used by replacing the low-frequency value $\beta_{\rm TF}$ 
with the high-frequency value $\beta_{\rm HF}=\sqrt{3/5}\vfermi$ 
~\cite{resonance_shifts_spill_out_hydro,PhysRevB.51.7497,electron_density_hydro_profile, first_GNOR_paper,Raza_2015, PhysRevB.60.15550}. 

Naturally, the question arises which characteristic velocity should be used, in general
~\cite{PhysRevB.51.7497,Aers1980NonradiativeSP}. 
A model within the continuum approach that interpolates between the two limits has 
been provided by Halevi
~\cite{PhysRevB.51.7497}. 
Specifically, Halevi considers the Euler-Drude model in $(\vec{r}, \omega)$-space and
introduces a frequency-dependent characteristic velocity $\halbeta$. Consequently,
Halevi's extension to the standard hydrodynamic model reads 
\begin{align}
 -i \omega \vec{J}(\vec{r}, \omega) &= \epsilon_0\volplasmafreq^2 \vec{E}(\vec{r}, \omega) 
                                       - \halbeta^2 \nabla \rho(\vec{r}, \omega) 
																			 - \gamma \vec{J}(\vec{r}, \omega). 
\label{eq:def_Halevi_Euler_Drude_equation_r_omega}
\end{align}  
 
The longitudinal dielectric function with wavenumber $k$ and frequency $\omega$ 
that is derived from 
Eqs.~(\ref{eq:continuity_eq_time_position}) and (\ref{eq:def_Halevi_Euler_Drude_equation_r_omega}) 
is subsequently expanded up to second order in the parameter $k\vfermi/\omega$. A 
comparison with a similar expansion within the Boltzmann-Mermin model is then 
performed. Halevi's approach, therefore, allows to isolate possible extensions 
of the Euler-Drude model without introducing higher orders of spatial nonlocality,
{
i.e. avoiding contributions that , e.g., scale with the second or higher powers 
of the Laplacian $\nabla^2$, or $\nabla (\nabla \cdot)$, or other combinations 
of vectorial differential operators. 
}
Further, Halevi's approach guarantees local charge conservation due to Mermin's 
relaxation time approximation
~\cite{PhysRevB.1.2362}. However, we would like to
stress that the high-frequency, collisionless limit of the Boltzmann-Mermin model, 
i.e., the Vlasov model
~\cite{PhysRevB.60.15550},
itself does not introduce any exchange-correlation effects in terms 
of pure electron-electron interaction
~\cite{universe7040108}. 

Upon incorporating in $\halbeta$ the dispersion introduced by the asymptotic 
expansion of the Boltzmann-Mermin model and comparing with the Bloch hydrodynamic 
model, Halevi obtains 
~\cite{PhysRevB.51.7497} 
\begin{align}
\halbeta^2 & = \frac{\frac{3}{5}\omega + \frac{i\gamma}{3}}{\omega + i \gamma}\vfermi^2.
\label{eq:halevi_beta_square}
\end{align}
The cross-over regime is characterized by the Drude collision rate $\gamma$. 
Indeed, Eq.~(\ref{eq:halevi_beta_square}) interpolates between the low-frequency velocity 
$\beta(\omega\ll\gamma)=\beta_{\rm TF}$ in the collision-dominated limit and the high-frequency 
velocity $\beta(\omega\gg\gamma)=\beta_{\rm HF}$ in the field-dominated limit
~\cite{PhysRevB.51.7497}. 
We refer to the system of equations (\ref{eq:continuity_eq_time_position}) and 
(\ref{eq:def_Euler_equation}) in the respective limits the low- and high-frequency 
Euler-Drude model.
For typical 
metals $\volplasmafreq$ lies in the visible or ultraviolet, such that 
$\gamma^2\ll\volplasmafreq^2$ and, therefore, the high-frequency value $\beta_{\rm HF}$ 
is preferred at optical frequencies $\omega\sim\volplasmafreq$. 
For intermediate frequencies, $\beta_{\rm H}$ exhibits 
dispersion which Halevi relates to a phase mismatch between pressure and density fluctuations 
~\cite{PhysRevB.51.7497}. 

In order to deepen the physical understanding of the transition from $\beta_{\rm TF}$ to 
$\beta_{\rm HF}$, we recall, as pointed out in Ref.~\onlinecite{universe7040108}, 
that the Halevi model may be viewed as the longitudinal projection of the viscoelastic model 
derived by Conti and Vignale. While Halevi's viewpoint focusses more on a formal correspondence 
to fluid dynamics, Conti and Vignale
~\cite{PhysRevB.60.7966} 
emphasize elastic contributions that are well-known from the elastic properties of solids. 
The latter dynamics introduces both, an elastic shear and a compressibility of the electron 
continuum. The latter remains constant over the  frequency range $\hbar\omega\ll E_{\rm F}$. 
In contrast, the hydrodynamic limit of the viscoelastic model is determined via the same 
compressibility augmented with a kinematic viscosity. The hydrodynamic bulk viscosity actually 
vanishes. This is similar to a hypothesis made by G.G. Stokes about the total pressure which 
is assumed to be independent of the temporal change of the fluid density during a uniform 
dilatation
~\cite{stokes_hypothesis}.

The range of applicability of the viscoelastic model is set by the contraints
~\cite{PhysRevB.60.7966,giuliani_vignale_2005}
\begin{align}
   k\ll 2k_{\rm F} \,\,\, , \,\,\, 
	 \omega \gg k \vfermi \,\,\, , \,\,\, 
	 \omega, \gamma \ll E_{\rm F} / \hbar.
\label{eq:scope_viscoelastic_model}
\end{align} 
Here, the first, third and fourth inequalities facilitate the interpretation of the 
underlying dynamics in terms of the Boltzmann equation with collisions, where the
momentum relaxes according to Mermin's recipe on the timescale $2\pi/\gamma$. The 
second inequality implies the transition to the viscoelastic paradigm confined by 
the hydrodynamic and elastic limit
~\cite{giuliani_vignale_2005}.

The formal equivalence of the latter continuum limits stems from the momentum 
conservation equation which can be rephraised in terms of the current density 
according to
\begin{align}
 -\omega(\omega + i\gamma) \vec{J}(\vec{r}, \omega)
    & = 
    -i\omega \volplasmafreq^2\epsilon_0\vec{E}(\vec{r}, \omega). \nonumber \\
    + \left[\tilde{\beta}^2 
		- \frac{4}{3}i\omega \tilde{\eta}\right]\nabla \left[\nabla \cdot \vec{J}(\vec{r}, \omega)\right] 
    &
		+i\omega \tilde{\eta}\nabla\times\nabla\times\vec{J}(\vec{r}, \omega).
\label{eq:viscoelastic_current_conserv_r_omega}
\end{align}
Here, in the high-frequency limit, the center-of-mass velocity in an infinitesimal 
volume around a given position originates from the temporal change of the displacement 
of an infinitesimal charge element due to compression and/or elastic shear
~\cite{PhysRevB.60.7966}.

When comparing this result with the longitudinal and transverse response of the 
Boltzmann-Mermin model up to the second order in the wavevector, it turns out
that the velocity $\tilde{\beta}$ is just 
the Thomas-Fermi velocity and that the parameter $\tilde{\eta}(\omega)$ provides 
an interpolation between the velocity of elastic shear waves $\betashear$ and 
kinematic viscosity $\eta (\omega)$ according to
\begin{align}
-i\omega\tilde{\eta}(\omega) &= \betashear^2(\omega) - i \omega \eta(\omega).
\label{eq:viscoelastic_shear} 
\end{align}
Here, the kinematic viscosity $\eta$ and the elastic shear velocity $\betashear$ 
are real quantities given by
\begin{align}
   \betashear^2(\omega) & = \frac{\omega^2}{\omega^2+\gamma^2}\frac{v^2_{\rm F}}{5}. \\
   \eta(\omega)         & = \frac{\gamma^2}{\omega^2+\gamma^2}\frac{v^2_{\rm F}}{5\gamma}.
\label{eq:viscoelastic_real_parameters} 
\end{align}
At this point, we would like to note that, similar to compression, also the shear 
is constrained by Pauli's exclusion principle as manifest by the appearance of the 
Fermi velocity in $\betashear$ and $\eta$. Apart from that, the material parameters 
are independent of position. 
Thus, within the bulk longitudinal (transverse) waves remain longitudinal (transverse). 
However, analogous to the theory of elasticity, at a material interface both polarisations 
mix 
~\cite{morse1953methods}. 
This represents a key property for plasmonic nano-particles, where the evanescent 
waves of surface plasmons can thus couple to bulk plasmons. For instance, transverse
polarized radiation impinging onto a cylinder may excite bulk plasmons
~\cite{RUPPIN2001205}. 

Considering that longitudinal quantities exhibit a vanishing curl, we can deduce
the connection  between the viscoelastic parameters and the Halevi velocity by 
combining Eq.~(\ref{eq:viscoelastic_current_conserv_r_omega}) with the continuity 
equation (see Eq.~(16) of 
~\cite{universe7040108}) to yield
\begin{align}
   \halbeta = \beta^2_{\rm TF} - \frac{4}{3}i\omega\tilde{\eta}(\omega).
\label{eq:halevi_vel_to_viscoelastic_shear}
\end{align}
Accordingly, the fact that the Halevi velocity $\beta_{\rm H}$ is a complex quantity
stems from the interpolation between the fluid- and solid-like response which, in turn,
is encoded in the asymptotic expansion of the Boltzmann-Mermin model. Conceptually, 
the additional dispersion (as compared to the Bloch model) originates completely in 
viscoelasticity
~\cite{universe7040108}. 
More precisely, within this perspective, the low-frequency, collision-dominated limit 
is viewed as the hydrodynamic regime and the high-frequency, collision-less limit is 
identified with the elastic regime
~\cite{PhysRevB.60.7966}. 
Since, we have for typical metals that $\volplasmafreq\gg\gamma$, the plasmonic response 
of metallic structures has to be regarded as being predominantly elastic. 
Therefore, and contrary to what is sometimes stated in the literature, the velocity 
$\beta_{\rm HF}$ is not entirely the result of the compressibility within the Thomas-Fermi
model. Instead, $\beta_{\rm HF}$ also features contributions from elastic shear.
  
Furthermore, the elastic response at short time scales ($2\pi/\omega\ll2\pi/\gamma$) and 
the viscous response at longer time scales ($2\pi/\omega\gg2\pi/\gamma$) are characteristic 
of highly viscous fluids
~\cite{PhysRevB.60.15550}, 
such as glycerin or resin
~\cite{landau1986theory_elasticity}. 
As proposed by Maxwell, the deformation of such fluids due to periodic, external forces 
induces internal shear stress, that is damped on a certain time scale $\tau$. As the period
$2\pi/\omega$ of the external forces progresses from values much lower than $\tau$ to 
values much greater values than $\tau$, the elastic response changes from solid- to 
(viscous-)liquid-like characteristics 
~\cite{landau1986theory_elasticity}. 
Since in the plasmonic case $\tau$ is roughly given by $2\pi/\gamma$, it is the rate of 
Drude collisions with the ionic background, that determines the extent of the solid-like 
response and, thus, of the continuum's tendency to restore its equilibrium shape. In the 
elastic regime, the periodic deformation is so fast, that collisions cannot relax the 
internal stress within one period. Note, again, that the magnitude of the shear is 
determined by the Fermi velocity [Eqs.~(\ref{eq:viscoelastic_shear}) and 
(\ref{eq:viscoelastic_real_parameters})], so that the internal stress is not only given 
by collisions but also tied to the Pauli principle. Since both paradigms are based on 
the continuum approximation we stick to the more general term \textit{electron continuum}. 

Interestingly, not only the longitudinal, but also the transverse dielectric function 
derived from Eq.~(\ref{eq:viscoelastic_current_conserv_r_omega}) is formally equivalent 
to a hydrodynamic dielectric function with a characteristic, squared transverse velocity 
$\beta^2_{\mathrm{T}}(\omega) = -i\omega\tilde{\eta}(\omega)$. 
In the elastic regime, we have  $\beta_{\mathrm{T}}\approx\beta_{\rm sh}(\omega\gg\gamma)=\vfermi/\sqrt{5}$. 
We emphasize this fact, since, for instance, Garc\'{i}a-Moliner and Flores
~\cite{GarciaMolinerFlores_nonspec_conduct_surfaces}
have applied the hydrodynamic dielectric function for the longitudinal \textit{and} the
transverse response of a metallic half-space in the limit $\omega\sim \omega_{\rm p}\gg \gamma$, 
with $\beta_{\rm L}=\beta_{\rm HF}$ and $\beta_{\rm T}=\beta_{\rm sh}(\omega\gg\gamma)$. 
Hence, the transverse and longitudinal dielectric function turn out to be high-frequency 
limits of the viscoelastic model -- corrections towards intermediate frequencies are thus 
at hand.
  
Having completed the discussion of charateristics of the Halevi model and obtaining a deeper 
understanding due to the viscoelastic interpretation, we now proceed, in the next section, to 
a first investigation of the model's impact on light-matter interaction within an analytical 
test case. 

\section{\label{sec:Ruppin_cyl_halevi}
         Extended Ruppin-Mie theory for a nonlocal, cylindrical scattering problem}

Analytical expressions for quantities that describe the light-matter interaction at metallic 
nanostructures are often tied to highly symmetric scatterers. An infinitely-extended circular
cylinder with radius $a$, homogeneous along its rotational axis and placed within vacuum
represents such an archetypal scatterer
~\cite{bohren2008absorption}. 
It further represents a very good approximation to realistic cylinders with high aspect ratios
~\cite{PhysRevB.95.115441}.
 
\begin{figure}
	\centering
	\includegraphics[scale=0.5]{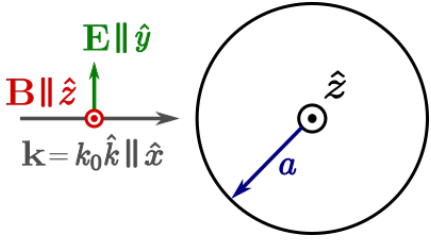}
	\caption{\label{fig:sketch_normally_excited_cylinder} 
	         Sketch of the infinitely-extended cylindrical scatterer excited normally 
					 with a plane wave propagating along the $x$-direction and polarized normal 
					 to the rotational axis ($\hat{z}$). 
					 The electromagnetic problem effectively reduces to two dimensions.
					 }
\end{figure}

Analytical formulas for the fields, given local and dispersive response within the metal, 
have been derived based on Mie-theory
~\cite{bohren2008absorption}
and have later been extended by Ruppin to allow for linear hydrodynamic reponse
~\cite{RUPPIN2001205}. 
In the latter work, the incident wave is injected perpendicular to the cylindrical axis
and the electric field is polarized also orthogonal to this axis, thus leading to an
effectively two-dimensional problem (cf. Fig.~\ref{fig:sketch_normally_excited_cylinder}).
Specifically, this setup allows for bulk modes in the metal 
~\cite{RUPPIN2001205}, as we have discussed in Sec.~\ref{sec:justification_halevi_model}.

Following a discussion of Melnyk and Harrison 
~\cite{PhysRevB.2.835},
Ruppin demands the absence of an infinitesimally thick surface charge density, which 
ultimately leads to a continuity of the radial component of the electric field. Since 
this is tied to the absence of the normal current right below the surface, similar to
the hardwall boundary condition
~\cite{resonance_shifts_spill_out_hydro,Raza_2015,PhysRevB.91.115416}, and the tangential 
current may, in principle, be finite in this region, this auxiliary boundary condiction
(ABC) is also known as slip boundary condition
~\cite{PhysRevB.84.121412}. 
Such a tangential surface current must not be mistaken with an infinitesimally thick 
surface current sheet, which would lead to a discontinuous, tangential magnetic field 
at the surface. In addition to the above, the continuity of the tangential electric 
field and normal magnetic flux field is enforced. 

The transverse response is encoded by the Drude dielectric function given by
\begin{align}
   \epsilon_{\rm Drude}(\omega) &= 1 - \frac{\omega^2_{\rm p}}{\omega(\omega+i\gamma)},
\label{eq:Drude_diel_func}
\end{align}
which gives for the wavenumber of the internal transverse mode
\begin{align}
   k^2_{\rm T} &= k^2_{0} \, \frac{\epsilon_{\rm D}(\omega)}{ \epsilon_{\rm BG}}.
\label{eq:transverse_wavenumber_drude}
\end{align}
Here, $k_0 = \epsilon_{\rm BG}\omega/c$ is the wave number of the incident wave with 
background dielectric function $\epsilon_{\rm BG}$. In the following, we use vacuum
as the background material and thus set $\epsilon_{\rm BG} = 1$.

In addition, the longitudinal response is given by the (high-frequency) Euler-Drude 
dielectric function 
\begin{align}
   \epsilon_{\rm L}(k, \omega) &= 1 - \frac{\omega^2_{\rm p}}{\omega(\omega+i\gamma) - \beta^2_{\rm HF}k^2}.
\label{eq:dielectric_function_high_freq_hydro}
\end{align} 
The corresponding wavenumber of the internal longitudinal mode derives from the 
implicit equation
\begin{align}
  \epsilon_{\rm L}(k_{\rm L}, \omega) &= 0 \,.
\label{eq:root_of_epsilon_longitudinal}
\end{align}
Completely analogous to standard Mie-theory, the incident, scattered and internal 
transverse portions of the electric field are expanded into solenoidal vector cylindrical 
harmonics. However, in the internal region an irrotational vector cylindrical harmonic is 
added to account for longitudinal waves. Solving the above-discussed set of boundary 
conditions for this Ansatz, Ruppin derives the multipole expansion coefficients of 
the scattered 
field as    
\begin{align}
   s_n & = -\frac{\left[c_n  + D_n(k_{\rm T} a) \right] J_n(k_{0}a) - 
	                \sqrt{\epsilon_T(\omega)}J'_n(k_{0}a)}{\left[c_n + D_n(k_{\rm T} a) \right] H_n(k_{0}a) - 
									                                       \sqrt{\epsilon_T(\omega)}H'_n(k_{0}a)},  
\label{eq:Ruppin-scatt-coeff}
\end{align}
where $\quad D_n(x) = J'_n(x) / J_n(x)$.

Here, $J_n(x)$ and $H_n(x)$ are, respectively, the Bessel and outgoing Hankel functions
of order $n$.\\
The nonlocal correction term $c_n$ is given by
\begin{align}
   c_n & = \frac{n^2}{k_{\rm L} a} \left[D_n(k_{\rm L} a) \right]^{-1}
	         \frac{\epsilon_T(\omega)-1}{k_0 a \sqrt{\epsilon_T(\omega)}}.
\label{eq:Ruppin-scatt-coeff_nonlocal_corr}
\end{align}
This nonlocal correction originates from the presence of the longitudinal field and
vanishes in the limit $\beta\to0$, where the longitudinal nonlocality is not resolved. 
Here $n$ counts the multipole order. 
For the monopole, $n=0$, we recall from Ref.~\onlinecite{bohren2008absorption}, 
that
\begin{align}
s_0 \propto (k_0a)^4\left( \sqrt{\epsilon_{\rm{T}}(\omega)} - 1\right),
\end{align}
in a local material description with vacuum as the host material, and provided that 
optical frequencies and radii within the range of $a\le10$nm are considered. Therefore, 
for a spatially local response, we do not expect a significant monopole contribution 
to the field. Further, the nonlocal coefficient $c_0$ vanishes and the same can be 
shown for the monopole coefficient of the internal, longitudinal field. In contrast, 
significant contributions are expected from the dipole ($n=1$) as well as the quadrupole 
($n=2$), for which the nonlocal coefficients $c_1$ and $c_2$ cannot be neglected.

From the discussion in Sec.~\ref{sec:justification_halevi_model}, we deduce that 
the Halevi model does not introduce any higher-order spatial derivatives as compared 
to the high-frequency Euler-Drude model. Otherwise, additional (spatial) boundary 
conditions would have to be introduced. We thus use the wavenumbers of the Halevi 
model within Eqs.~(\ref{eq:Ruppin-scatt-coeff}) and (\ref{eq:Ruppin-scatt-coeff_nonlocal_corr}).
\begin{figure*} 
 \includegraphics[scale=0.5]{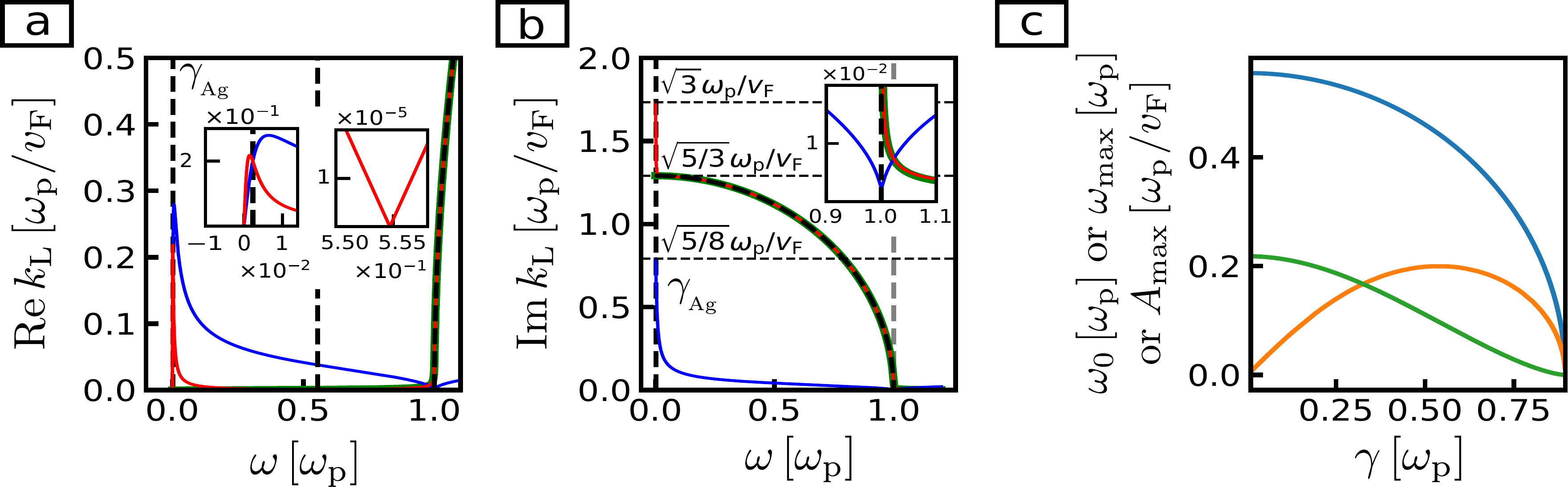}
  
 \caption{\label{fig:halevi_wavenumber} 
          Illustration of the frequency-dependence of the longitudinal wavenumber 
					$k_{\rm L}$ for the high-frequency Euler ($\gamma=0$; black dashed line) 
					and Euler-Drude 
					(green solid line), the Halevi model (red solid line) and GNOR model (blue 
					solid line) for $\gamma = 0.021\text{eV}$ (silver). 
					For details on the latter model, see the discussion in Sec.~\ref{sec:comparison_Halevi_GNOR}.
					The vertical dashed lines mark the damping rate and kink frequency (black) 
					as well as the volume plasma frequency for silver (grey). 
					Panel [a]: Real part of the wavenumber with 
										 insets showing a new local maximum (left) and kink (right). 
				  Panel [b]: Imaginary part of the wavenumber with an inset displaying how 
					           the Halevi and Euler-Drude model soften the transition at the 
										 plasma frequency. 
				  Panel [c]: Kink frequency $\omega_0$ (blue line) as well as the frequency 
					           $\omega_{\rm{max}}$ of the local maximum (orange line) and its
										 amplitude $A_{\rm max} = \operatorname{Re}k_{\rm L}(\omega_{\rm max})$ 
										 (green line) in the range $0.01<\gamma / \volplasmafreq < \sqrt{4/5}$. 
				  } 	
\end{figure*}

As Ruppin points out, the longitudinal wavenumber describes the propagation of the 
longitudinal modes. Therefore, we will now determine its features within the Halevi 
model and deduce important differences compared to the high-frequency Euler(-Drude) 
model. To do so, we start with the remark, that Eqs.~(\ref{eq:halevi_beta_square}),
(\ref{eq:dielectric_function_high_freq_hydro}) and (\ref{eq:root_of_epsilon_longitudinal}) 
actually only provide the square of the wavenumber given by
\begin{align}
   k^2_{\rm L} &= \frac{\omega(\omega + i \gamma) - \volplasmafreq^2}{\halbeta^2},
\label{eq:longitudinal_wavenumber_square_Halevi}
\end{align}
where we substituted $\beta_{\rm HF}\mapsto\halbeta$. 
This disguises some of the wavenumber's characteristics in the complex plane. We fill 
in the blanks by sticking to a passive system, thus enforcing a non-negative imaginary 
part. By choosing the principal branch of the complex square root, the real part must
remain positive.  

In Fig.~\ref{fig:halevi_wavenumber}, we depict the frequency dependence of the longitudinal
wavevector for different material models of silver (see Tab.~\ref{tab:metal_material_params_and_damping} 
for material parameters). 
From this, we can infer several features which distinguish the Halevi from the Euler(-Drude) model. 
Specifically, below the volume plasma frequency a kink occurs (cf. right inset of 
Fig.~\ref{fig:halevi_wavenumber}a). 
The actual frequency is derived as follows: To first order in $\gamma/\omega_{\rm p}$, 
the real part of the squared wavenumber changes sign at $\omega_{\rm p}$ - being negative 
for frequency values below $\omega_{\rm p}$. 
At the same time, the imaginary part changes 
sign at the real frequency
\begin{align}
   \omega_0 = \volplasmafreq \sqrt{\frac{4}{13} - \frac{5}{13}\frac{\gamma^2}{\volplasmafreq^2}}.
\end{align}
as long as $0 < \gamma < \sqrt{4/5} \omega_{\rm p} \approx 0.894\volplasmafreq$, where  
$\omega_0\approx0.555\volplasmafreq$ for typical metals (cf. Fig.~\ref{fig:halevi_wavenumber}c). 
Thus, the real part of the longitudinal wavenumber, given by
\begin{align}
\Re{k_{\rm L}}(\omega) = \sqrt{\frac{1}{2}\left( \Re{k^2_{\rm L}}(\omega) + \sqrt{\left[\Re{k^2_{\rm L}}(\omega)\right]^2 + \left[\Im {k^2_{\rm L}}(\omega)\right]^2}\right)},
\label{eq:Re_k_L_halevi}
\end{align}
should exhibit a root at $\omega_0$. 

At lower frequencies, we observe a local maximum (see left inset of 
Fig.~\ref{fig:halevi_wavenumber}a). Different real parts 
in the different models should, in principle, lead to spatial phase 
differences for a given frequency. Interestingly, we observe in 
Fig.~\ref{fig:halevi_wavenumber}c, that the frequency 
of this local maximum intially increases with $\gamma$ until a value 
of roughly $0.2\omega_{\rm p}$ is reached. 
Beyond this value, the frequency of the local maximum drops and vanishes 
at $\gamma = \sqrt{4/5}\omega_{\rm p}$ along with the corresponding amplitude.  

Regarding the imaginary part of the longitudinal wavenumber, we notice from
Fig.~\ref{fig:halevi_wavenumber}b, that a peak occurs at 
low frequencies. This is due to the small velocity of the low-frequency density 
waves, which, in the Halevi model, reaches the Thomas-Fermi value of 
$\beta_{\rm TF}=\vfermi/\sqrt{3}<\beta_{\rm HF}$. 
Increasing the ratio $\gamma/\volplasmafreq$ leads to a broadening of the 
peak and for the plasma frequency of bulk silver, this peak extends into the 
infrared frequency range. Following Ruppin, this is tantamount to a faster
decay of the longitudinal modes 
~\cite{RUPPIN2001205} 
in this frequency range. Still looking at the silver curves, we deduce from 
the inset of Fig.~\ref{fig:halevi_wavenumber}b, that for 
$\omega> \volplasmafreq$, the decay is approximately given by the Drude damping 
due to the ionic background.  

Now that we have determined a number of bulk features introduced by the Halevi 
model, we proceed to considering the surface resonances of nonlocal cylinders. 
In particular, we focus on the quasi-static approximation, i.e., where the 
incident wavelength and the skin depth are much larger than the cylinder radius.
In this case, the phase of the plane wave is approximately constant across the 
cylinder's cross section
~\cite{maier2007plasmonics}. 
Since we expect that the resonances are at typical wavelengths, i.e. of the order
of a few $100$nm, the quasi-static approximation is roughly valid for radii 
below a few $10$nm. As derived in the Appendix 
the leading order contribution of the nonlocality manifests itself in the 
dispersion relation according to
\begin{align}
   \omega_n & \approx \omega_{\rm sp} - \frac{i\gamma}{2} 
	                    + \frac{\beta_{\rm HF}}{2} \frac{n}{a} 
											  \left(1 - \frac{2i}{9} \frac{\gamma}{\omega_{\rm sp}}\right),
\label{eq:halevi_slip_cyl_sp_resonance_approx}
\end{align}
for $n>0$. The first three terms on the r.h.s. recover the result of the linearized Euler-Drude 
model regarding the nonlocal blueshift. 

\begin{table}
	 \caption{\label{tab:metal_material_params_and_damping} 
	          List of the material parameters for several metals. 
		 		  	The Fermi velocity for Au and Ag is determined from the plasma frequency using 
			  	  the bare electron mass. The Drude rates of Cu, Al and Zn were calculated using 
					  $\gamma= 2\pi/\tau$ with the relaxation time $\tau$ at $77$K taken from 
					  ~\cite{ashcroft1976solid}. 
					  From these values, based on Eq.~(\ref{eq:imag_part_halevi_cylinder_surface_resonance}), 
					  the ratio $\Delta_1$ of the nonlocal ($\Im \omega_1 - \gamma/2$) to the local 
					  damping ($\gamma/2$) as well as the nonlocal damping amplitude $A_1$, according 
					  to Eq.~(\ref{eq:A_factor_halevi_cyl_SP_damping}), are infered for $a=1$nm.}
	 	\begin{ruledtabular}
	 		\setlength{\tabcolsep}{6pt}
	 		\begin{tabular}{ld{3}d{3}d{3}d{3}d{3}}
	 			& \multicolumn{1}{c}{$\omega_{\rm p}~[\text{eV}]$} & \multicolumn{1}{c}{$\gamma~[\text{eV}]$} & \multicolumn{1}{c}{$v_{\rm F}~[10^6\frac{\text{m}}{\text{s}}]$} & \multicolumn{1}{c}{$\Delta_1$} &
	 			\multicolumn{1}{c}{$A_1 [10^{-4}]$} \\
	 			\hline
	 			Ag~\text{\cite{PhysRevB.6.4370}} &9.143&~~~0.021&1.407&0.012&2.828\\
	 			Au~\text{\cite{PhysRevB.6.4370}}&8.846&~~~0.059&1.376&0.012&  8.070\\	 	
	 			Zn~\text{\cite{ashcroft1976solid}}&10.143~\text{\cite{gross2018festkoerperphysik}}&~~~0.172&1.508&0.012&20.678\\	 	
	 			Al~\text{\cite{ashcroft1976solid}}&15.363~\text{\cite{gross2018festkoerperphysik}}&~~~0.064&1.989&0.010&5.041\\	 	
	 			Cu~\text{\cite{ashcroft1976solid}}&7.504~\text{\cite{gross2018festkoerperphysik}}&~~~0.020&1.233&0.013&3.195\\
	 		\end{tabular}
	 	\end{ruledtabular}
\end{table}

\begin{figure*}
	\centering
	\includegraphics[scale=0.31]{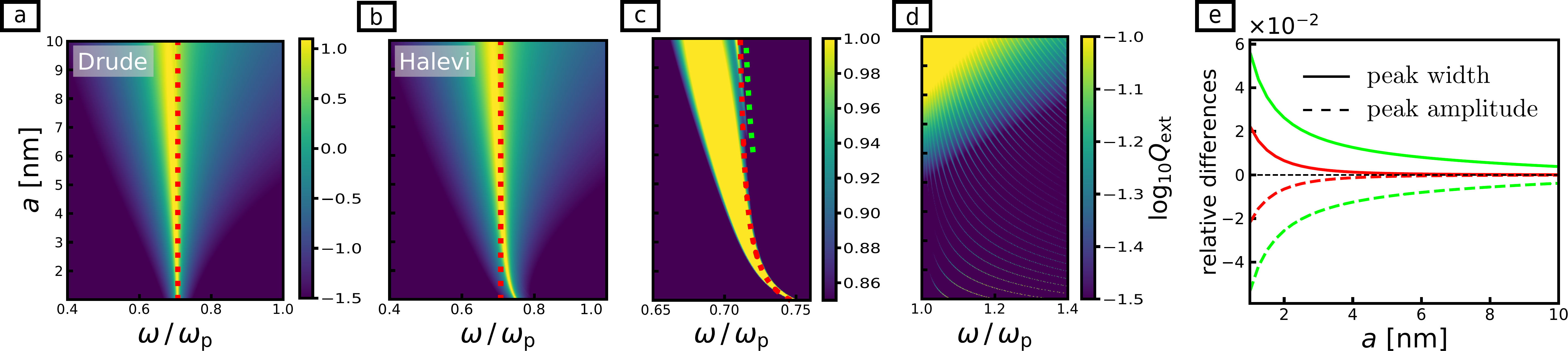}
	\caption{\label{fig:extinction_efficiencies}	 
		Illustration of the extinction efficiencies of an infinite circular silver cylinder situated
		in vacuum and excited by a plane wave propagating in a plane perpendicular to the cylinder 
		axis with polarization perpendicular to this axis. 
		Panel [a]: Color-coded representation of the extinction efficiency as a function of radius 
		and frequency modelled using the Drude model. The red dotted vertical line 
		represents the local, electrostatic surface plasmon resonance. 
		Panel [b]: Same as panel [a], but substituting the Drude with the Halevi model. The colorbar 
	  is that of Panel [a].										 
		Panel [c]: Blow-up of the dipolar and quadrupolar surface plasmon peaks present in  
	  panel [b]. The real part of Eq.~(\ref{eq:halevi_slip_cyl_sp_resonance_approx})  
		is plotted for the dipole (red dotted line) and quadrupole (green dotted line) 
		terms. 									 
		Panel [d]: 	
		Extinction efficiency for the Halevi model in the region beyond the volume plasma frequency exhibiting 
		a nonlocal sequence of volume plasma resonances being blue-shifted towards lower radii 
		and overlapping with the dipole surface resonance at larger radii. 
		Panel [e]: Radius dependence of the relative difference of the Halevi and the Euler-Drude 
		model for high frequencies for the widths (solid lines) and amplitudes 
		(dashed lines) of the dipole (red) and quadrupole (green) peak. These values 
		have been obtained by a Lorentzian fit in the interval $0<\omega<\omega_{\rm p}$ 
		using the Levenberg-Marquardt algorithm and only the sum element of the 
		extinction efficiency belonging to the respective resonance. A cutoff at 
		$\omega = \omega_{\rm p}$ prevents additional fitting errors due to the 
		excitation of bulk plasmons.
	} 
\end{figure*}

The imaginary part of the surface resonance frequency for a multipole of order $n$ is given by
\begin{align}
   \Im \omega_n = - \frac{\gamma}{2} \left(1 + \frac{2}{9} \frac{\beta_{\rm HF}n}{\omega_{\rm sp}a}\right).
\label{eq:imag_part_halevi_cylinder_surface_resonance}
\end{align}
It is proportional to the collision rate and negative, which conforms with our passivity
requirement. In contrast to the well-known result for the high-frequency Euler-Drude model, 
the damping incorporates a nonlocal contribution. Such a feature has already been observed 
by Halevi in the dispersion relation of the volume plasmon 
~\cite{PhysRevB.51.7497}. 
Halevi refers to this additional damping channel as 'collision-modified Landau damping'. 
While, the plasmonic frequencies lie deep within the elastic regime, the novel damping term 
vanishes for $\gamma/\volplasmafreq=0$. Hence, the viscous damping, also proposed in Refs.
~\onlinecite{Jewsbury_1981,viscoel_eps_near_zero_thin_films}, 
which contributes in the intermediate, viscoelastic regime, is essential for this damping 
channel to be present. 

At this point, we wish to emphasize, that the calculations in the Appendix~\ref{app-sec:cylindrical_surface_plasmons}
are based on the limiting case $\vfermi n / \volplasmafreq a\ll1$ so that the nonlocal 
damping contribution is small compared to the Drude damping. 
In Tab.~\ref{tab:metal_material_params_and_damping}, we provide the relevant parameters
for a number of metals that are used in plasmonics. From this, we infer that the nonlocal 
damping of the dipole for cylinder radius $a=1$nm only contributes with about a percent 
of the local Drude damping, with Copper having the largest value, so that the aforementioned
assumption is well justified.

Analyzing the dissipation for frequencies close to $\omega_n$, Eq.~(\ref{eq:imag_part_halevi_cylinder_surface_resonance})
appears to suggest a description of the system in terms of a Drude dielectric function  
where the Drude damping rate is amended with a term that scales with the inverse of the 
cylinder radius. Such an approach would allow to avoid the intricacies that come along 
with the nonlocal hydrodynamic description matched by a slip boundary condition.
Nonetheless, while such an approach would, indeed, yield the same expression for the 
imaginary part of the dispersion relation, we abstain from such an identification as
this would blur the clear conceptual distinction between bulk and surface dynamics 
established by the bulk dielectric function and the set of surface boundary conditions, 
respectively. 
Nevertheless, we would like to note that such an interpretation via a size-dependent damping 
rate is reminiscent of a proposition made in Refs.
~\onlinecite{PhysRevB.48.18178,limitation_elec_mean_free_path} 
which phenomenologically describes the measured extinction of small spherical silver 
nano-particles immersed in different host media by limited-mean-free path effects. In 
these works, the thus corrected damping rate was formally written as
\begin{align}
   \gamma(a) &= \gamma + A \frac{\vfermi}{a}, 
\end{align} 
and utilized within a bulk Drude dielectric function. 
The value of the corresponding linewidth parameter
~\cite{PERSSON1993153} 
$A$ is then determined for different embedding scenarios by a fit to measured extinction 
data and the difference in values is motivated by a discussion of possible surface damping 
mechanisms.
Within the Halevi model and when we assume that the cylinder is embedded in vacuum, 
we derive the linewidth parameter as
\begin{align}
   A = \frac{n}{9}\frac{\beta_{\rm HF}}{v_{\rm F}} \frac{\gamma}{\surfplasmafreq}.
\label{eq:A_factor_halevi_cyl_SP_damping}
\end{align}
We note, that its structure is not only determined by the actual geometry and the choice 
of boundary condition but also depends strongly on the actual bulk response. Within the 
high-frequency Euler-Drude model equipped with the same boundary condition, this novel 
effect would be completely absent. While in 
Refs.~\onlinecite{PhysRevB.48.18178,limitation_elec_mean_free_path}, 
the phenomenological surface damping has been used to describe the dipole mode for 
spherical scatterers only, we would like to point out that our analysis implies that a
similar structure is to be expected for higher multipole modes, 
too.
{ 
For results of the impact of leading-order nonlocality on higher-order surface (and 
bulk) plasmons excited in metal spheres, we refer the reader to 
Ref.~\onlinecite{doi:10.1021/nn406153k}. 
}
Further, other scatterer geometries (for which numerical simulation might be required)
will exhibit different values of the linewidth parameters, even for dipoles. 
Specifically, for silver spheres in vacuum a value of $A_{\rm Ag}=0.025$ for the dipole
resonance has been determined experimentally
~\cite{limitation_elec_mean_free_path}. 
A glance at Tab.~\ref{tab:metal_material_params_and_damping} reveals, that the $A$-value 
for the dipolar resonance of silver nano-wires as derived from the Halevi model in conjunction 
with the slip boundary condition is about two orders of magnitude smaller. Owing to a larger
value of $\gamma/\omega_{\rm sp}$, the corresponding linewidth parameter for zinc nano-wires 
is about one order of magnitude larger.
To complete our analysis of the nonlocal damping term, we briefly mention, that the 
application of the so-called GNOR model to metallic spheres with the very same ABC 
has also led to a motivation of the bespoken limited mean-free-path effects 
~\cite{first_GNOR_paper,Mortensen_meso_edyn_at_surf_2021,Svendsen_role_diff_surf_scatt}. 
In Sec.~\ref{sec:comparison_Halevi_GNOR}, we will provide a comparison of the Halevi and 
this alternative model.
 
We conclude the this section by stating that the additional damping should lead to a 
broadening of the quasi-static resonance peaks while, at the same time, decreasing the 
amplitude. This should be visible in the extinction efficiency defined as
~\cite{RUPPIN2001205,doi:10.1021/nn406153k}
\begin{align}
   Q_{\rm ext} &= - \frac{1}{2 k_0a} \sum\limits_{n=-\infty}^{\infty} \Re s_n.
\label{eq:extinc_efficiency_general_p_pol_cyl}
\end{align} 
In Fig.~\ref{fig:extinction_efficiencies}, we depict the extinction efficiency as a 
function of both, radius and frequency, for the Drude {(panel a)} and Halevi model {(panel b)}. The results for
the high-frequency Euler-Drude model are not shown because they are very similar to those
of the Halevi model. We observe, that all models display a strong dipole and a weaker 
quadrupole surface plasmon mode. The peak of the latter branches off at about $6$nm. 
For smaller radii, this mode is hardly discernable, since the right flank of the dipole 
mode is of a much larger magnitude. The nonlocality introduces bulk plasmons beyond the 
volume plasma frequency -- a feature that is absent in the Drude model and is clearly
visible in {Fig.~\ref{fig:extinction_efficiencies}(d)}. 
For smaller radii, the bulk and surface resonances of the Halevi model are blue-shifted. 
In order to investigate in more detail the dipolar and quadrupolar surface resonances, 
we provide in {Fig.~\ref{fig:extinction_efficiencies}(c)} a corresponding blow-up, 
where the real part of Eq.~(\ref{eq:halevi_slip_cyl_sp_resonance_approx}) 
is displayed with dotted lines.  
The dipole follows this approximation quite well up to radii of about $4$nm beyond 
which the approximation quickly tends to the retarded quadrupole resonance. 

In fact, the retarded dipole falls below the local, electrostatic surface plasma frequency 
$\omega_{\rm p}/\sqrt{2}$, which is not expected by Eq.~(\ref{eq:halevi_slip_cyl_sp_resonance_approx}). 
This turns out to be a shortcoming of the quasi-static approximation, which relies on 
the smallness of the parameter $a \omega_{\rm sp}/c$, an effect that would be pronounced 
for a surrounding medium with $n_{\rm{BG}}>1$, e.g., glass. For larger radii, the concomittant 
retardation yields a redshift of the modes, which is already visible in the Drude model 
-- otherwise the dipolar and quadrupolar mode would, in this model, be degenerate. The 
dispersion relation Eq.~(\ref{eq:halevi_slip_cyl_sp_resonance_approx}), applied to the Halevi 
model, further introduces the parameter $\beta_{\rm HF}/a \omega_{\rm sp}$, leading to 
a small blue shift, which is larger for the quadrupole resonance. While the nonlocal 
blueshift decreases with increasing radius, the retardation-based redshift increases.  
Further, when considering both, retardation and the leading order in the longitudinal 
nonlocality, products of powers of these parameters may arise. In particular, the 
product of the parameters themselves is proportional to 
$\beta_{\rm HF}/c\sim10^{-2}$, such a correction is comparable to the nonlocal parameter.
According to {Fig.~\ref{fig:extinction_efficiencies}(c)} the overall blueshift of the 
quadrupole still appears to surpass any redshifting contributions for radii of about 
$a\approx10$nm. 
The retardation redshift has been phenomenologically described for small metallic 
spheres by employing a local material response within the framework of a self-
consistent treatment of Mie theory
~\cite{maier2007plasmonics}. 

Further, we would like to note that beyond radii of $5.5$nm the right flank of the 
dipolar peak, which exhibits a growing asymmetry towards larger radii, increasingly
overlaps with the region of bulk plasmons. It is, thus, conceivable that when deduced 
from numerical simulations, the field distributions of the latter become modulated 
by the surface plasmon mode. 

In order to quantify the effect of  nonlocal damping, we fit the peaks in both, the 
high-frequency Euler-Drude as well as the Halevi model with Lorentzians and deduce 
both, the width and amplitude as functions of the cylinder radius and depict the
results in {Fig.~\ref{fig:extinction_efficiencies}(e)}. 
The difference in widths relative to the value of the Euler-Drude model increases 
towards lower radii and the difference is always larger for the quadrupole mode. 
Both are expected from Eq.~(\ref{eq:imag_part_halevi_cylinder_surface_resonance}). 
Further, the increase in width is accompanied by a decrease in amplitude, which, 
again, turns out to be more pronounced for the quadrupole mode. We would like to
note that the combined contributions due to retardation and nonlocality may be 
more pronounced for larger radii.

\section{\label{sec:deriv_drift_diffusion_current}Translation into time domain}

In section Sec.~\ref{sec:Ruppin_cyl_halevi}, our considerations have been carried
out within the frequency domain. Further insight can be obtained from time-domain
considerations. For this purpose, we consider Eqs.~(\ref{eq:def_Halevi_Euler_Drude_equation_r_omega}) 
and (\ref{eq:halevi_beta_square}). After a few algebraic manipulations, we find
\begin{align}
      \left(-i \omega + \gamma\right)\vec{J}(\vec r, \omega) 
	 &= 
	    \epsilon_0 \volplasmafreq^2 \vec E (\vec r, \omega) 
	    - \beta^2_{\rm HF} \nabla \rho(\vec r, \omega) \nonumber \\ 
   & \quad  
	   + \frac{4v^2_{\rm F}}{15} \frac{\gamma}{\gamma - i\omega} \nabla \rho(\vec r, \omega).  
\label{eq:halevi_model_current_conservation}
\end{align} 
From our previous discussions, we recall that the Halevi model introduces additional 
dispersion  in the nonlocal gradient term of the high-frequency Euler-Drude equation, 
which leads to the last term on the r.h.s. of Eq.~(\ref{eq:halevi_model_current_conservation}). 
In order to incorporate dispersive response into a time-domain numerical framework, 
the technique of auxiliary differential equation (ADE) is often applied 
~\cite{lpor_DGTD_review}. 
In our specific case, we equate the last term of Eq.~(\ref{eq:halevi_model_current_conservation}) 
with a current $\vec{J}_{\rm D}$ multiplied by the negative damping rate $-\gamma$. 
Performing an inverse Fourier transform to the time-domain yields the ADE
\begin{align}
    \partial_t \vec{J}_{\rm D}(\vec r, t) + \gamma \vec{J}_{\rm D}(\vec r, t) 
		&= 
		- \frac{4 v^2_{\rm F}}{15} \nabla \rho(\vec r, t). 
\label{eq:drift_diffusion_ADE}
\end{align} 
which, apart from the temporal-derivative term on the l.h.s., is reminiscent of 
Fick's first law, describing an ordinary diffusion current (hence, the label $\rm{D}$).
The Halevi equation [Eq.~(\ref{eq:halevi_model_current_conservation})] can then 
be rephrased in the time domain as 
\begin{align}
   \partial_t \vec{J}(\vec{r}, t) 
	 =& 
	 \, \epsilon_0\volplasmafreq^2\vec{E}(\vec{r}, t) 
	 - \beta^2_{\rm HF} \nabla\rho(\vec{r}, t) \nonumber \\   
   &  
	 - \gamma \left[\vec{J}(\vec{r}, t)+\vec{J}_{\rm D}(\vec{r}, t)\right]. 
\label{eq:current_conservation_Halevi_drift_diffusion}
\end{align}

In order to clarify the origin of the auxiliary current, we consider the viscoelastic 
counterpart of Eq.~(\ref{eq:halevi_model_current_conservation}), which can be derived 
from Eq.~(\ref{eq:viscoelastic_current_conserv_r_omega}) by applying the continuity 
equation as well as the Gau\ss-Maxwell law and considering the longitudinal current 
only. The identification with the Halevi velocity is then mediated by 
Eq.~(\ref{eq:halevi_vel_to_viscoelastic_shear}), which can be recast into
\begin{align}
   \halbeta^2 &= \beta^2_{\rm HF} + \frac{4}{3} \left[-i \omega\tilde{\eta}(\omega) - \frac{\vfermi^2}{5} \right].
\label{eq:Halevi_vel_additional_dispersion_viscoelastic_identification}
\end{align}
The squared velocity $\vfermi^2/5= -i\omega\tilde{\eta}(\omega)|_{\omega\gg\gamma}$ 
corresponds to purely elastic waves and serves to restore the Thomas-Fermi velocity in 
the hydrodynamic limit. A comparison of Eq.~(\ref{eq:Halevi_vel_additional_dispersion_viscoelastic_identification}) 
with the prefactors of the density-gradient terms in Eq.~(\ref{eq:halevi_model_current_conservation}) 
eventually allows for an identification of the additional dispersion with viscoelastic 
shear, encoded in $\tilde{\eta}$. As apparent in the second term on the r.h.s. of 
Eq.~(\ref{eq:viscoelastic_current_conserv_r_omega}), $\tilde{\eta}$ enters with a 
numerical prefactor of $4/3$. Multiplying the latter with $\vfermi^2/5$, the squared 
diffusive velocity $v^2_{\rm D} = 4v^2_{\rm F}/15$ is recovered. 

While the dynamics of $\mathbf{J}$ is coupled directly to that of $\mathbf{J}_{\rm{D}}$, 
as apparent in Eq.~(\ref{eq:current_conservation_Halevi_drift_diffusion}), the dynamics 
of the novel current also depends on that of $\mathbf{J}$ via the longitudinal stress 
$\nabla \rho(\vec r, t)$ emerging in Eq.~(\ref{eq:drift_diffusion_ADE}).
We can obtain insight into the modified dynamics, introduced via the auxiliary current, 
by inspecting the induced charge density. For this reason, we introduce the decomposition  
\begin{align}
   \rho    &= \tilde{\rho} + \rho_{\rm D}.
\label{eq:charge_density_decomposition}
\end{align}

Using Eq.~(\ref{eq:charge_density_decomposition}) in the continuity equation, we 
obtain $\nabla \cdot \vec{J}_{\rm D} = - \partial_t \rho_{\rm D}$ and rephrase 
Eq.~(\ref{eq:drift_diffusion_ADE}) as
\begin{align}
   v^2_{\rm D} \triangle \tilde{\rho} 
	 &= 
	 \partial_t^2 \rho_{\rm D} 
	 + \gamma \partial_t \rho_{\rm D} 
	 - v^2_{\rm D} \triangle \rho_{\rm D}.
\label{eq:ADE_in_terms_of_charges}
\end{align}
Equating the r.h.s. to zero yields the Cattaneo equation
~\cite{Compte_1997}.

Back in 1948, Cattaneo modified the ordinary diffusion current of a classical 
substance, given by Fick's first law, by employing a relaxation-time approach. 
This enforces a finite propagation velocity of the classically diffusing quantity 
~\cite{Compte_1997}. 
Quantities that obey this type of equation perform a hybrid diffusive-wavelike 
propagation
~\cite{PhysRev.131.2013,morse1953methods}. 
Which propagation paradigm dominates, depends on the relative temporal change 
of the respective quantity
~\cite{PhysRev.131.2013} 
defined by
\begin{align}
   f_{\rm D} &= \frac{1}{2\pi} \left|\frac{\partial_t \rho_{\rm D}}{\rho_{\rm D}}\right|.
\label{eq:Chester_rate_for_rho_D}
\end{align}
in our case. Comparing this rate to $\gamma$, the r.h.s. of Eq.~(\ref{eq:ADE_in_terms_of_charges}) 
suggests wavelike transport in the collision-less regime ($f_{\rm D}\gg\gamma$) 
and diffusive transport in the collission-dominated regime. Accordingly, close 
to the collision-less regime, a small diffusive contribution obstructs a purely 
wave-like motion
~\cite{morse1953methods}. 
However, we would like to note that this hybrid propagation is modulated by the 
dynamics of $\tilde{\rho}$, which, in general, cannot be treated as a source term.

Actually, both types of propagation are found for the total induced charge. In
order to facilitate the discussion, we introduce the rate $f$ similar to 
Eq.~(\ref{eq:Chester_rate_for_rho_D}), but in terms of $\rho$. Considering 
Eqs.~(\ref{eq:drift_diffusion_ADE}) and (\ref{eq:current_conservation_Halevi_drift_diffusion}) 
in conjunction with the Gau\ss-Maxwell law and the continuity equation, yields 
\begin{align}
   - {\volplasmafreq^2} \rho 
	 = 
	 \left({\gamma}\partial_t 
	 - \beta^2_{\rm TF}\nabla^2\right)\rho,
\label{eq:total_induced_charge_dynamics_halevi_diffusive_regime}
\end{align}
in the hydrodynamic ($f,f_{\rm D}\ll\gamma$) and
\begin{align}
   - \volplasmafreq^2 \rho 
	 = 
	 \left(\partial^2_t 
	 - \beta^2_{\rm HF}\nabla^2\right)\rho,
\label{eq:total_induced_charge_dynamics_halevi_wave_regime}
\end{align}
in the elastic regime ($f,f_{\rm D}\gg\gamma$). Accordingly, in these regimes the 
conduction electrons exhibit diffusive and wave-like transport, respectively, with 
a propagation velocity that changes from $\beta_{\rm TF}$ to $\beta_{\rm HF}$. This 
is linked to the insight, that the bare Euler-Drude model yields a Cattaneo-type 
differential equation for the totally induced charge. 

However, we would like to note that the Cattaneo-type dynamics of the Euler-Drude model 
is further modulated by the Coulomb interaction which introduces a
restoring force that keeps the electrons within the metal (see, for instance, the 
l.h.s. of Eqs.(\ref{eq:total_induced_charge_dynamics_halevi_diffusive_regime}) and 
(\ref{eq:total_induced_charge_dynamics_halevi_wave_regime})). 
Incidentally, since  the restoring term originates in the electric field that enters 
Eq.~(\ref{eq:halevi_model_current_conservation}) and since screened dielectric 
functions are constructed as those functions which connect the current not only to 
some external but also to the internal field, we find 
\begin{align}
   \epsilon_{\rm H}(k, \omega) 
	 & =  1 - \frac{\volplasmafreq^2}{\omega(\omega + i \gamma) 
	        - \beta^2_{\rm HF}k^2 + \gamma v^2_{\rm D}k^2 / (\gamma - i\omega) }.
\label{eq:halevi_model_longitudinal_dielectric_fctn}
\end{align}
Accordingly, the restoring term, proportional to $\volplasmafreq^2$, only enters 
through the numerator of the second term on the r.h.s. of Eq. (\ref{eq:halevi_model_longitudinal_dielectric_fctn}). 
In the hydrodynamic regime, the pole of this very term turns purely diffusive, i.e. 
$\omega=-i(\beta^2_{\rm TF}/\gamma)k^2$, and becomes purely wavelike in the elastic
regime, i.e. $\omega^2=\beta^2_{\rm HF}k^2$.

Finally, we would like to note that the quantum-mechanical nature of the conduction 
electrons enters the equations of motion solely via the characteristic velocities, 
which are proportional to the Fermi velocity.
Apart from this, these equations are similar to those of classical material dynamics. 
This can be understood by reformulating the fourth condition of the viscoelastic model,
(see Eq.~(\ref{eq:scope_viscoelastic_model})) to
\begin{align}
   \ell \sim \vfermi / \gamma \gg 1 / k_{\rm F} \sim \lambda_{\rm F}.
\end{align}
As a result, the (bulk) mean-free-path $\ell$ is much larger than the Fermi wavelength 
$\lambda_{\rm F}$ within the viscoelastic model. Hence, a representative electron 
conducts a random walk in which successive collisional events are independent of each 
other and quantum interference effects are absent
~\cite{giuliani_vignale_2005}. 

\section{\label{sec:comparison_Halevi_GNOR}Comparison with the GNOR model}

In the previous Sec.~\ref{sec:deriv_drift_diffusion_current}, we have shown, that the 
Halevi model extends the high-frequency Euler-Drude model via an additional current 
that shares some similarities with a diffusive current according to Fick's first law. 
An extension of the Euler-Drude model due to diffusive dynamics is also a key element 
of the GNOR model
~\cite{first_GNOR_paper}. 
Therefore, it is worthwile, to discuss similarities and differences of the GNOR and 
the Halevi model.  

In Refs.
~\onlinecite{first_GNOR_paper,Raza_2015,bozhevolnyi2016quantum,Mortensen_meso_edyn_at_surf_2021} 
a generalized formal approach to nonlocality within linear response theory is discussed. 
Given a position ${r}$ within the electron continuum, as a central approximation, the 
influence of nonlocal processes is limited to a neighborhood with an extent that is much
smaller than the length scale of variation of the electric field. Within the constitutive 
equations, this allows for a Taylor expansion of the field of the electron continuum 
where each term corresponds to a certain order of nonlocality. Given a homogeneous, 
isotropic medium with local response described by the Drude model, the leading-order-nonlocal 
current density is then given by
\begin{align}  
  -i\omega \vec{J}(\vec{r}, \omega) 
	&= 
	\epsilon_0 \volplasmafreq^2\vec{E}(\vec{r}, \omega) 
	- \gamma \vec{J}(\vec{r}, \omega) \nonumber  \\ 
	& 
	- \nu^2 \nabla \left[ \epsilon_0\nabla \cdot \vec{E}(\vec{r}, \omega)\right].
\label{eq:general_nonlocal_longi_current_with_xi_in_r_omega}
\end{align}
The last term on the r.h.s. lumps together the effects of the considered order of nonlocality 
that extends the local Drude response. 

Here, we would like to note that the differential operator in space is motivated by 
the Euler-Drude equation. As remarked in 
Ref.~\onlinecite{Raza_2015}, the nonlocality of 
the latter only manifests itself in the longitudinal current reponse. Since it has been 
pointed out, that surface plasmons are affected not only by the longitudinal response
~\cite{bozhevolnyi2016quantum}, 
we add that the transverse nonlocality of the viscoelastic model, 
Eq.~(\ref{eq:viscoelastic_current_conserv_r_omega}), serves as a plausible amendment.

Eventually, in order to construct the GNOR model, first the center-of-mass current 
$\rho_0 \vec{v}$ is defined to fulfill the high-frequency Euler-Drude model, while 
the total current is defined by the sum of the latter and a diffusive current
~\cite{first_GNOR_paper,Raza_2015,bozhevolnyi2016quantum}, 
such that
\begin{align}
   \vec{J}(\vec{r}, \omega) 
	 &= 
	 \rho_0 \vec{v}(\vec{r}, \omega) - D \nabla \rho(\vec{r}, \omega).
\label{eq:convection_diffusion_current}
\end{align}
Accordingly, the diffusion is basically introduced by hand, following Fick's first 
law
~\cite{bozhevolnyi2016quantum,Compte_1997}, 
given a diffusion constant $D$. In order to prohibit the build-up of charges, this 
constant has to be positive.

It can then be shown, that
\begin{align}
   \nu^2 
	 &= 
	 \beta^2_{\rm HF} + D(\gamma - i \omega) . 
\label{eq:GNOR_compression_diffusion_xi}
\end{align}
We would like to note that the r.h.s. of Eq.~(\ref{eq:GNOR_compression_diffusion_xi}) 
provides the counterpart of Halevi's interpolation formula defined in Eqs.~(\ref{eq:halevi_beta_square}) 
and (\ref{eq:Halevi_vel_additional_dispersion_viscoelastic_identification}). While both 
models include the high-frequency Euler-Drude model, which introduces a nonlocal process 
with real-valued length scale $\ell_{\beta}=\beta_{\rm HF}/\omega$, both, the GNOR and
the Halevi model, 
Eqs.~(\ref{eq:GNOR_compression_diffusion_xi}) and (\ref{eq:Halevi_vel_additional_dispersion_viscoelastic_identification}),
respectively, provide an extension of $\beta_{\rm HF}$ to complex-valued velocities.

\begin{figure*}
	 \includegraphics[scale=0.65]{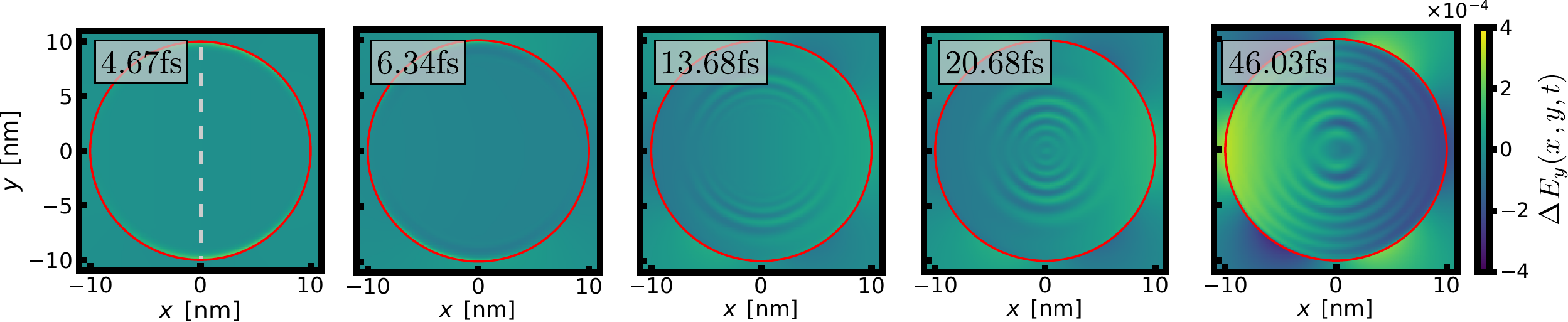}
	 \includegraphics[scale=0.65]{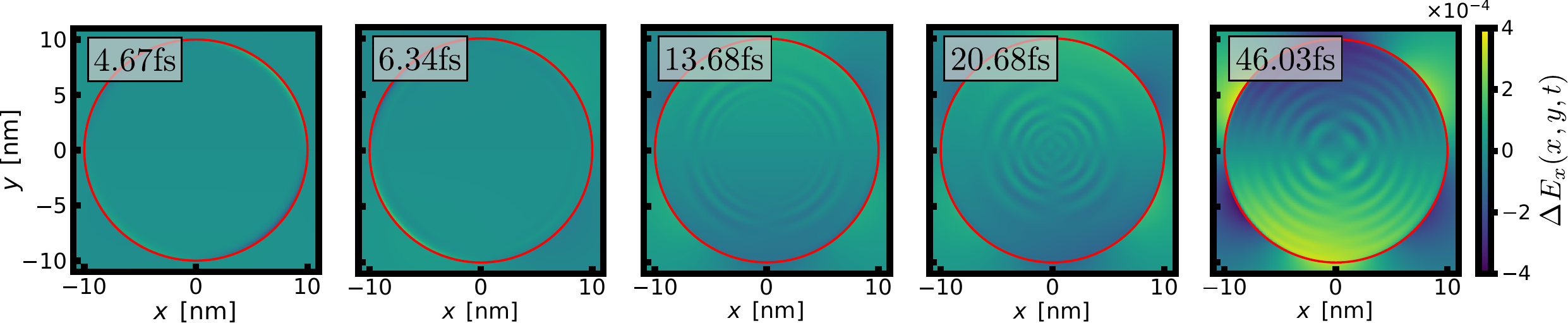}
	 \caption{\label{fig:rel_diff_E_y_linHd_halevi} 
	          Illustration of the time evolution of the relative difference of $E_y$ (upper row) 
						and $E_x$ (lower row) given by Eq.~(\ref{eq:def_rel_diff_linHd_vs_halevi}) between 
						the Halevi and high-frequency Euler-Drude model across the horizontal cross section of a cylindrical
						scatterer situated in vacuum. The cylinder has a radius of $10$nm and 
						is made of silver with material parameters according to Tab.~\ref{tab:metal_material_params_and_damping}
						(see also Fig.~\ref{fig:sketch_normally_excited_cylinder} for the scattering setup). 
						Spatially confined field oscillations build up and propagate towards the center 
						until they eventually spread across the entire disk. The vertical dashed line in the 
						upper row marks the line considered in Fig.~\ref{fig:rel_diff_E_y_center_cut}.
						}	
\end{figure*}

In the original GNOR model, the diffusion constant is estimated via $D\sim \vfermi^2/\gamma$ 
~\cite{first_GNOR_paper}, 
yielding a diffusive length scale $\ell_{\rm D}=\sqrt{D/\omega}\sim \vfermi/\sqrt{\gamma\omega_{\rm p}}$ 
which we have evalutated at plasmonic frequencies. 
Therefore, considering typical metals this diffusive scale is much larger than 
$\ell_{\beta}\sim \sqrt{\gamma/\volplasmafreq} \ell_{\rm D}$. 
In contrast, within the Halevi model $\ell_{\beta}$ surpasses the diffusive length scale, the 
latter being given by $\ell^{\rm Hal}_{\rm D}\sim\sqrt{\gamma/\volplasmafreq}(\vfermi/\volplasmafreq) 
                       \sim
											 \sqrt{\gamma/\volplasmafreq}\ell_{\beta}$. 
The mismatch in the frequency dependence of each extension to $\beta_{\rm HF}$ has also been 
noted in Ref.~\onlinecite{viscoel_eps_near_zero_thin_films}.

Further, within the GNOR model, the additional current reads
\begin{align}
   \gamma \vec{J}_{\rm GNOR}(\vec{r}, \omega) & = \left[\nu^2 - \beta^2_{\rm HF}\right] \nabla \rho(\vec{r}, \omega ),
\label{eq:def_GNOR_aux_eq_r_omega}
\end{align} 
where we isolated the part proportional to $\beta_{\rm HF}$ stemming from the Euler-Drude 
model. Upon inserting Eq.~(\ref{eq:GNOR_compression_diffusion_xi}) as well as the continuity 
equation, where the total current now obeys Eq.~(\ref{eq:convection_diffusion_current}) 
per construction, we find 
\begin{align}
   \gamma \vec{J}_{\rm GNOR}(\vec{r}, \omega) 
	  &= 
		D\gamma \nabla \rho(\vec{r}, \omega) 
		- D \nabla \left[ \nabla \cdot \vec{J}(\vec{r}, \omega)\right].
\label{eq:GNOR_aux_eq_r_omega}
\end{align}
The dynamics of the total current within the GNOR model then obeys
\begin{align}
-i\omega \vec{J}(\vec{r}, \omega) 
    &= 
		\epsilon_0 \volplasmafreq^2\vec{E}(\vec{r}, \omega) 
		- \gamma (\vec{r}, \omega) \vec{J}(\vec{r}, \omega) \nonumber \\
    &  
		- (\beta^2_{\rm HF} + D\gamma) \nabla \rho(\vec{r}, \omega) 
		+ D \nabla \left(\nabla \cdot \vec{J}(\vec{r}, \omega)\right).
\label{eq:GNOR_current_eq_with_aux_current_terms}
\end{align}
As a result, we see that the first term on the r.h.s. of Eq.~(\ref{eq:GNOR_aux_eq_r_omega}) 
adds to the density gradient term of the high-frequency Euler-Drude equation. Beyond that, 
the second term on the r.h.s. of Eq.~(\ref{eq:GNOR_aux_eq_r_omega}) formally reproduces 
the longitudinal shear contribution of Eq.~(\ref{eq:viscoelastic_current_conserv_r_omega}). 
The corresponding coefficient has to be positive, which is consistent with $D>0$. If we 
then utilize the frequency-dependent (real-valued) diffusion constant deduced in Ref.~\onlinecite{Raza_2015} 
by equating the imaginary parts of Eqs.~(\ref{eq:GNOR_compression_diffusion_xi}) and 
(\ref{eq:halevi_beta_square}) given by
\begin{align}
   D & = \frac{v^2_{\rm D}}{\gamma}\frac{\gamma^2}{\omega^2+\gamma^2},
\label{eq:GNOR_D_from_imag_of_nu_and_halevibeta}
\end{align}
we see that the density gradient term of Eq.~(\ref{eq:GNOR_current_eq_with_aux_current_terms}) 
is equipped with the characteristic velocity $\beta^2_{\rm HF}$ in the high-frequency regime
and it is equipped with $\beta^2_{\rm HF}+v^2_{\rm D}> \beta^2_{\rm TF}$ in the low-frequency 
regime. This results from the fact that $D(\omega\gg\gamma)\sim \gamma/\omega^2$ and 
$D(\omega\ll\gamma)\sim 1/\gamma$, which is not the case in the Halevi model. 

Beyond that, by equating the real parts of Eqs.~(\ref{eq:GNOR_compression_diffusion_xi}) and 
(\ref{eq:halevi_beta_square}), we find the negative of the r.h.s. of Eq.~(\ref{eq:GNOR_D_from_imag_of_nu_and_halevibeta}) 
if we assume a real-valued diffusion constant.

Another difference between the GNOR and the Halevi model arises, when we derive the counterpart 
of Eq.~(\ref{eq:convection_diffusion_current}) for the Halevi model. For this purpose, we 
consider Eq.~(\ref{eq:halevi_model_current_conservation}) and define the total current as 
\begin{align}
   \vec{J}(\vec{r}, \omega) &= \rho_0 \vec{v}(\vec{r},\omega) + f_{\rm H}(\vec{r},\omega).
\label{eq:ansatz_for_total_Halevi_current}
\end{align}
Trying to restore the high-frequency Drude-Euler equation for $\rho_0\mathbf{v}$, we 
require the equivalence between $f_{\rm H}(\vec{r},\omega)$ and the residual terms and obtain
\begin{align}
   f_{\rm H}(\vec{r}, \omega) &= -\frac{v^2_{\rm{D}}}{\gamma} \frac{\gamma^2}{(\omega+i\gamma)^2} \nabla \rho(\vec{r}, \omega). 
\label{eq:correction_to_total_current_by_Halevi}
\end{align}
This additional current is not simply defined by Fick's first law. The difference originates 
in the way how the local equilibration is implemented in the Mermin approach. To yield 
equivalence with the Ansatz in Eq.~(\ref{eq:convection_diffusion_current}), the diffusion constant 
does not only have to be frequency-dependent, but also complex-valued. 

\begin{figure*}
	 \includegraphics[scale=0.65]{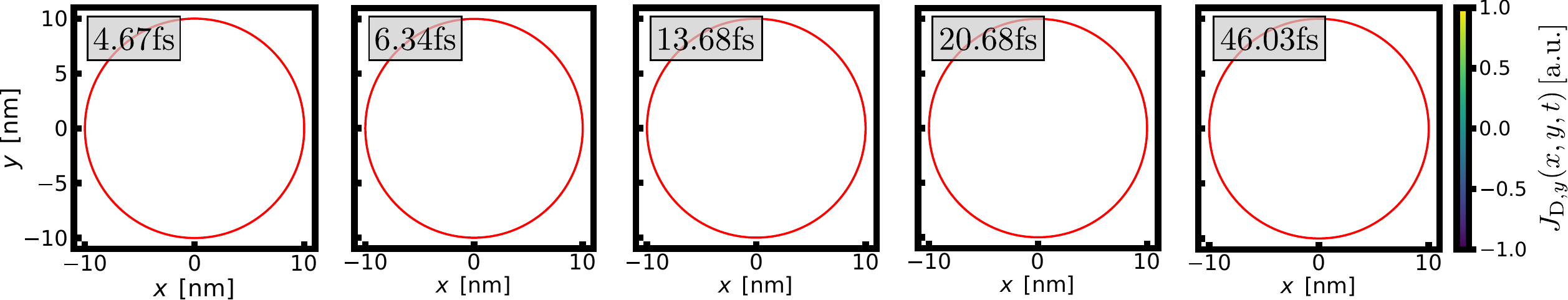}
	 \includegraphics[scale=0.65]{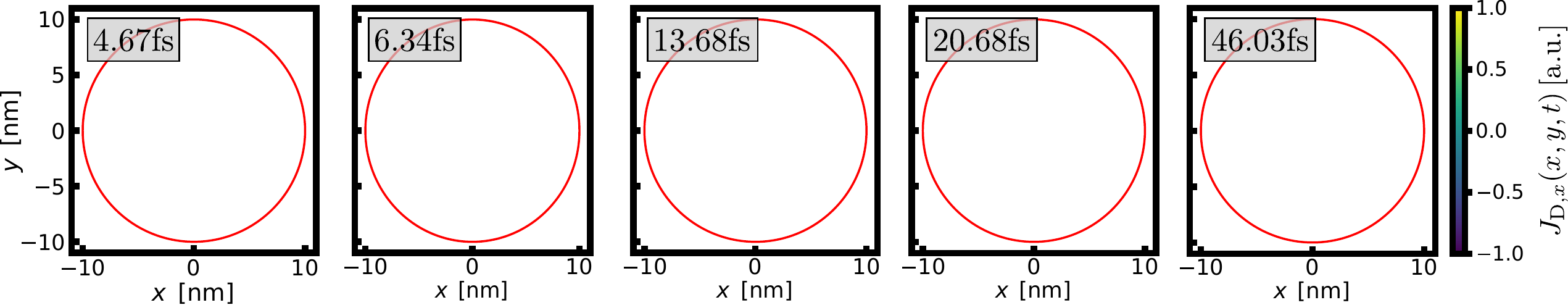}	
	 \caption{\label{fig:drift_current_x_and_y_comp} 
	                 Illustration of the the time evolution of the current components 
									 $J_{\text{D},y}$ (upper row) and $J_{\text{D},x}$ (lower row) of 
									 the auxiliary current [Eq.~(\ref{eq:drift_diffusion_ADE})] 
									 within the horizontal cross section of a silver cylinder (radius $a=10$nm,
									 see Fig.~\ref{fig:sketch_normally_excited_cylinder} for the general scattering 
									 setup). 
								   Starting from the surface region an oscillatory contribution propagates towards 
									 the cylinder's center and finally extends across the entire cross section.
									 }	
\end{figure*}

The particular choice of diffusive paradigm also influences the longitudinal wavenumber. 
Therefore, we added the results of the GNOR model to Figs.~\ref{fig:halevi_wavenumber}a 
and \ref{fig:halevi_wavenumber}b, using the approximation $D\approx \frac{\vfermi}{\gamma}$ 
that has been suggested in 
Ref.~\onlinecite{first_GNOR_paper}. 
First, the real part exhibits a low-frequency local maximum, just as the Halevi model. However, 
there is no additional kink root below the plasma frequency. Further, beyond the plasma frequency
the high-frequency Euler(-Drude)-model is not followed as closely. Considering the imaginary 
part, the amplitude of the low-frequency maximum is smaller than in the high-frequency Euler-Drude 
model as well as the Halevi model, owing to the larger velocity 
$\sqrt{\beta^2_{\rm HF}+D\gamma}>\beta_{\rm HF},\beta_{\rm TF}$. 
This larger velocity is also tied to a larger blueshift for resonances that potentially occur
within this frequency regime in nano-objects when considering a 
hardwall boundary condition. A similar observation has been made in
Ref.~\onlinecite{Svendsen_role_diff_surf_scatt}. 
Further, for frequencies beyond the volume plasma frequency, the GNOR model does not follow 
the high-frequency Euler-Drude model as closely as does the Halevi model. Specifically, the 
imaginary part increases in the GNOR-model instead of decreasing in the Euler-Drude and the
Halevi model.

We deduce, that a direct identification of both diffusive extensions of the high-frequency 
Euler-Drude model has to be excercised with caution although these models share certain 
qualitative similarities. Interestingly, $\vec{J}_{\rm D}$, when added to the shear-corrected 
pressure term ($\propto \beta^2_{\rm HF}$) yields a viscoelastic single-relaxation-time 
approximation of the diffusion current derived in Eq.~(31) of Ref.~\onlinecite{PhysRev.117.1252}. 
In this reference, a total current is derived, which is close to Eq.~(\ref{eq:convection_diffusion_current}), 
provided a real-valued diffusion constant. However, the derivation is confined to a steady-state 
case. The diffusion is related to the inhomogeneity of the Thomas-Fermi pressure and, therefore, 
$D\gamma = \vfermi^2/3$. This provides the nonlocal correction to Ohm's law. 
This correction is required to demonstrate, that within a metal a vanishing total current 
does not yield a vanishing electric field (as otherwise, complete screening of charges would 
always follow immediately).

Recently, it has been reported that the GNOR model leaves the scope of the (bulk) Boltzmann(-Mermin) 
model by introducing high-frequency (optical) diffusion
~\cite{Mortensen_meso_edyn_at_surf_2021}. 
Further, a microscopic justification of the diffusive extension inherent in the GNOR model 
has been proposed by resorting to surface response functions that employ the Feibelman 
$d$-parameter
~\cite{Svendsen_role_diff_surf_scatt, Mortensen_meso_edyn_at_surf_2021,PhysRevLett.118.157402}. 
Accordingly, to conclude this section, we would like to note, that the Halevi model provides 
an alternative to the GNOR model as a unifying description of the size-dependent frequency-shift 
and broadening of plasmonic modes in nanometer-structured metals which stays within the scope
of the Boltzmann-Mermin model. As a result, the Halevi model is firmly routed in bulk arguments. 
However, owing to the different scaling with the ratio $\gamma / \volplasmafreq$ (given a 
frequency-independent $D$), the nonlocal damping of the Halevi model is smaller in amplitude
relative to the GNOR model.

Finally, we would like to note that 
Ref.~\onlinecite{Svendsen_role_diff_surf_scatt} 
only considers the Boltzmann equation without the Mermin correction while the latter has been 
mentioned in one of the original publications on the GNOR model in 
Ref.~\onlinecite{Raza_2015}. 
Actually, in 
Ref.~\onlinecite{Kittel1963},
the Mermin correction has been linked to an additional, diffusive current. We stress this 
fact, since the Mermin-Ansatz for the single-relaxation-time approximation introduces an 
additional term which 
prohibits the occurrence of a charge sink in the continuity equation, as shown in 
Refs.~\onlinecite{PhysRevB.1.2362,wegner2020remarks}.

\section{\label{sec:halevi_phase_shift_simulations}Temporal evolution of fields in the Halevi model}

Eqs.~(\ref{eq:continuity_eq_time_position}), (\ref{eq:drift_diffusion_ADE}) and 
(\ref{eq:current_conservation_Halevi_drift_diffusion}) together with the slip boundary 
condition comprise the complete system of equations for Halevi model of plasmonic 
materials. 

In order to solve the Maxwell equations with this material model for an arbitrary 
geometry of a scatterer, we implement the time-domain version of the Halevi model 
via ADEs (see Sec.~\ref{sec:deriv_drift_diffusion_current}) into our home-made Discontinuous Galerkin Time-Domain 
(DGTD) approach
~\cite{lpor_DGTD_review}
which is a finite-element method that is specifically designed to solve equations 
in conservation form and we have utilized the algorithm in nodal form developed 
by Hesthaven and Warburton 
~\cite{hesthaven2007nodal}
and typically employ third-order Lagrange polynomials as basis functions and confirm
that fourth-order polynomials give the same results. 
Further, within the DGTD approach a numerical flux is introduced in order to couple adjacent 
elements. We utilize a pure upwind flux for the Maxwell equations 
~\cite{hesthaven2007nodal,lpor_DGTD_review} 
and employ for Eqs.~(\ref{eq:continuity_eq_time_position}) and 
(\ref{eq:current_conservation_Halevi_drift_diffusion}) the Lax-Friedrichs flux 
~\cite{hesthaven2007nodal}. 
Finally, we solve the ADE, Eq.~(\ref{eq:drift_diffusion_ADE}), using a central flux. The 
resulting DGTD spatial discretization yields a set of ordinary differential equations 
of first order in time which we solve via a 4th-order Low-Storage Runge-Kutta method 
with 14 stages 
~\cite{NIEGEMANN2012364}. 
The Drude and Euler-Drude model have been implemented 
within  DGTD and for details, we refer to 
Refs.~\onlinecite{lpor_DGTD_review} 
and~\onlinecite{Hille2016},
respectively.

For our subsequent simulations, we consider a cylindrical silver wire of radius $10$nm
(see Tab.~\ref{tab:metal_material_params_and_damping} for the material parameters)
in vacuum. The wire is illuminated by an electromagnetic plane wave with wave vector 
and electric field normal to the cylinder axis, i.e. identical to the analytical case 
of Sec.~\ref{sec:Ruppin_cyl_halevi} (see also Fig.~\ref{fig:sketch_normally_excited_cylinder}
for an illustraton of the setup). 

The exciting pulse is centered around $\omega_0 = 7.4$eV (which is roughly $0.8$ times 
the plasma frequency) and has a Gaussian envelope with a FWHM of $1.57$fs. This pulse
is injected into the system using the TF/SF technique
~\cite{lpor_DGTD_review}. 
Since the problem effectively reduces to two dimensions, the TF/SF contour is chosen 
to be a square with edge length $40$nm centered on the cylinder axis. The pulse is launched
from the left edge of the TF/SF contour so that the maximum at the TF/SF contour 
occurs at $t_0 = 4.67$fs. The scattered field box is bounded by the TF/SF contour 
and a centered square of edge length $440$nm. To prevent 
unphysical back reflexions, we surround the entire computational domain with perfectly-matched 
layers and apply Silver-M\"{u}ller boundary conditions at the outer boundary given by
a centered square with edge length $1.04\mu$m. 
The corresponding mesh is generated using Gmsh 
~\cite{gmsh_article}. 
In order to adequately resolve the cylindrical material interface, we utilize a minimal 
(maximal) insphere radius of $0.04$nm ($0.32$nm) for the finite elements near this
interface. 

\begin{figure*}
	 \includegraphics[scale=0.475]{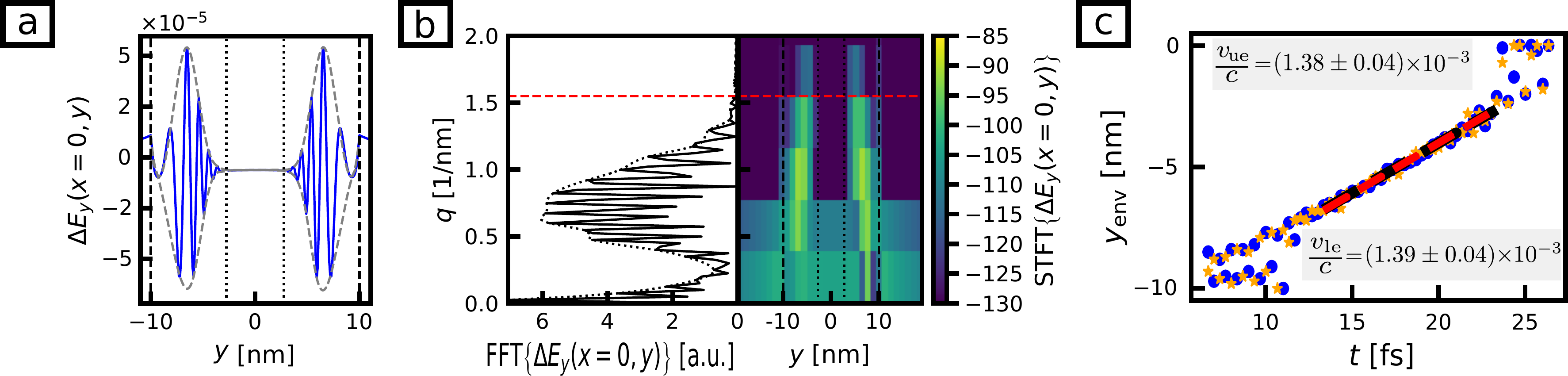}
	 \caption{\label{fig:rel_diff_E_y_center_cut} 
	          Snapshot of the relative difference of $E_y$ given by 
						Eq.~(\ref{eq:def_rel_diff_linHd_vs_halevi}) between the high-frequency 
						Euler-Drude and the Halevi model at time $t\approx13.68$fs along a vertical 
						through the cylinder, as outlined by the dashed line in 
						Fig.~\ref{fig:rel_diff_E_y_linHd_halevi}. 
						Panel [a]: Relative difference in ${E}_y$, displaying two wave pakets 
					  Panel [b]: Onesided, spatial FFT of the signal of Panel [a] as 
											 function of spatial frequency $q$ (left part), as well as a spectrogram 
											 exhibiting the Short-Time-Fourier-Transform (STFT) of the signal 
											 in Panel [a] using a Hamming window (right part). Panel [a] and 
											 the right part of panel [b] feature
											 the same $y$-axis. 
											 The spatial frequency resolution of the spectrogram is chosen 
											 as roughly $1\text{nm}^{-1}$ to bring out the positive chirp towards 
											 the center within the two wave pakets of panel [a]. The sampling 
											 rate of roughly $1\text{nm}^{-1}$ yields a Nyqvist limit of 
											 $q_{\text{max}}\approx 5 \text{nm}^{-1}$. 
											 The colorbar of the spectrogram ist truncated to yield a better 
											 contrast in the dominant spatial frequency range of roughly 
											 $q\in[0.25,1.5]\text{nm}^{-1}$ -- the upper bound being marked 
											 by a horizontal dashed line. 
					  Panel [c]: Position $y_{\rm env}$ of the global maximum (blue dots) and 
						           global minimum (orange stars) of the left wave paket (as 
											 exemplified in (a) for $t\approx13.68$fs) as a function of 
											 time. A linear fit is provided yielding estimates for the 
											 upper envelope (ue) and lower envelope (le) group velocity.}	
\end{figure*}

In Sec.~{\ref{sec:justification_halevi_model}}, we have elaborated on the Halevi model 
as an extension of the high-frequency Euler-Drude model. Therefore, we now proceed to
perform a direct comparison of the numerical results for the two models and focus on 
the relative difference of the spatio-temporal electric field distributions according 
to 
\begin{align}
   \Delta E_i(\vec{r}, t) 
	  &= 
		\frac{E_i^{\rm linHd}(\vec{r}, t)- E_i^{\rm Hal}(\vec{r}, t)}{\operatorname{max}_{(\vec{r}, t)} E_i^{\rm linHd}(\vec{r}, t)},
\label{eq:def_rel_diff_linHd_vs_halevi}
\end{align}
where $i\in\{x,y\}$, i.e., for the $x$- and $y$-component of the electric field. We would
like to recall that our excitation pulse propagates along the $x$-axis and is polarized 
along the $y$-axis. 

From Fig.~\ref{fig:rel_diff_E_y_linHd_halevi}, we infer that the $x$- and $y$-components 
exhibit differences in amplitude, which at $t\approx4.67$fs is most significant at the 
surface. As the pulse progresses, the relative differences assume the form of spatially 
confined oscillations close to the surface at $t\approx 6.34$fs. Around this time, the 
left flank of the pulses leaves the cylinder. Around $t\approx 13.68$fs, the oscillations 
have travelled about half the way to the cylinder's center and reach the center around 
$t\approx 20.68$fs. 
Subsequently, the oscillations start to disperse across the entire cross section of the
cylinder and, ultimately, become damped out. We provide the last time frame at $t\approx 46.03$fs 
in order to demonstrate that the oscillatory pattern has distributed over the cylindrical 
cross section. Incidentally, the relative difference at this time instance is reminiscent 
of the mode picture of hydrodynamic bulk plasmons 
~\cite{Hille2016}.
As such, the corresponding $y$-component posseses a mirror symmetry with respect to the 
$xz$-plane and is mapped onto itself by a rotation of $180$ degree around the cylinder
axis. The $x$-component, however, is mapped onto itself by a $90$ degree rotation. These 
symmetries are also present at the other time instances. 
Nonetheless, the differences in the field distributions are rather weak. 

In Sec.~\ref{sec:deriv_drift_diffusion_current}, we have shown, that the novel dispersion 
of the Halevi model (as compared to the high-frequency Euler-Drude model, Eq.~(\ref{eq:halevi_model_current_conservation})) 
translates into an additional current, that follows Eq.~(\ref{eq:drift_diffusion_ADE}). 
Accordingly, we expect, that the difference in the electric field is tied to the spatio-temporal 
evolution of this additional current. Consequently, in Fig.~\ref{fig:drift_current_x_and_y_comp} 
we display the $x$- and $y$-component of the additional current $\vec{J}_{\text{D}}$ at the 
same time instances where we depicted the field differences in Fig.~\ref{fig:rel_diff_E_y_linHd_halevi}. 
We observe that, at all time instances, both components concentrate the maximal amplitude 
at the surface of the cylinder. However, a small contribution is given by the spatially 
confined oscillations, which propagate in essentially the same way as the relative difference 
in the field components, also respecting the afore-discussed symmetry properties.

\begin{figure*}
	 \includegraphics[scale=0.225]{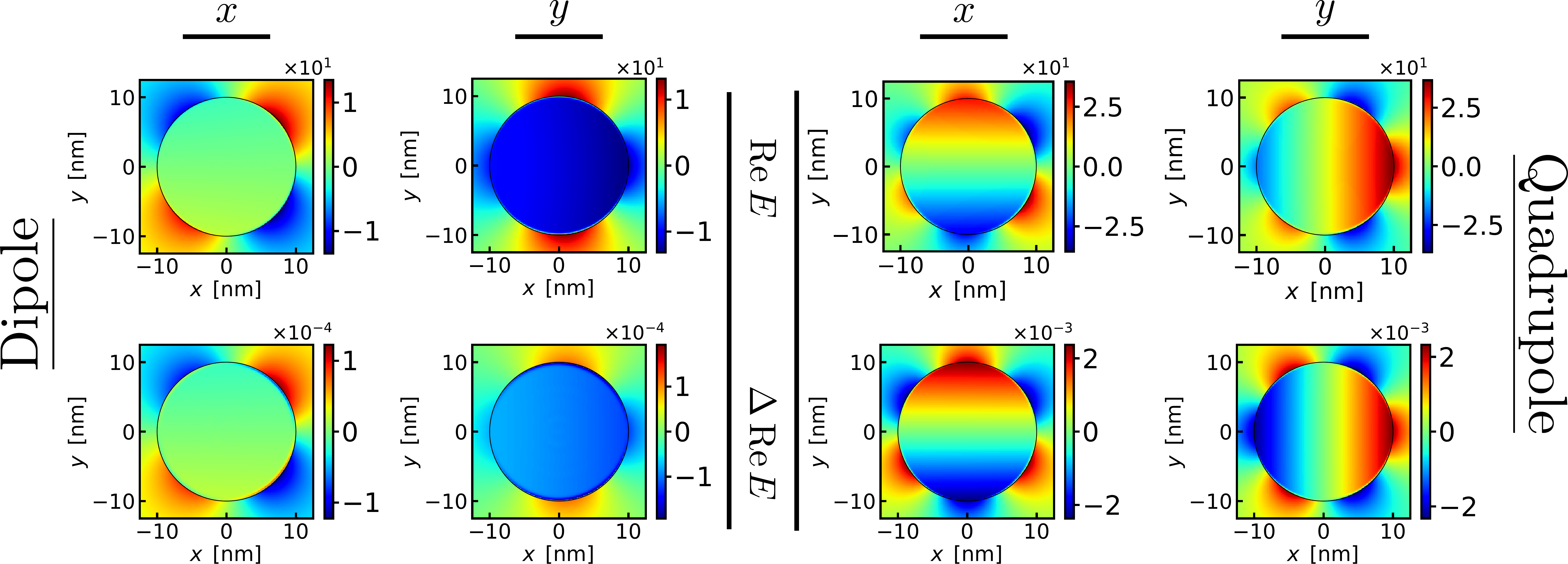}
	 \caption{\label{fig:mode_pictures_driftHD_and_relative} 
	          Illustration of the electric field distributions of dipolar (left panel) and 
						quadrupolar (right panel) resonances within the Halevi model (upper row) as 
						well as their relative differences to the corresponding modal distributions 
						for the high-frequency Euler-Drude model (lower row). The difference has been 
						normalized to the maximum value obtained with the Euler-Drude model at the
						respective frequency. 
						Positive (negative) values of the respective distribution are complemented by 
						positive (negative) values in the relative difference.  
					  The maximum enhancement reaches about a tenth of a promille at the dipole frequency and 
					  about 2 promille at the quadrupole frequency. 
            The relative difference inside the scatterer is  
					  more inhomogeneous for the quadrupolar mode.
						}
\end{figure*}

In order to measure, how fast the oscillations in the relative field difference propagate 
into the scatterer, we consider a cut through the cylinder center as indicated in 
Fig.~\ref{fig:rel_diff_E_y_linHd_halevi}. Given the pulsed nature of our excitation,
these oscillations build up and assume the form of wave pakets -- due to symmetry 
one in each half of the cut. In Fig.~\ref{fig:rel_diff_E_y_center_cut}(a) we provide 
an illustrative time frame. To highlight the spatial confinement, we determine the 
interpolative (upper and lower) envelope. 
Specifically, we sample the maxima and minima and perform corresponding cubic spline 
fits
~\cite{Yang_signal_theo_approach_envelope_ana}. 

This approach works well for positions away from the surface but experiences some
problems near the surface where the boundary conditions obstruct the formation of
well-developed wave paket. From the global extrema of the upper and lower envelopes, 
we can deduce the movement of the wave pakets which ultimately yields two estimates 
of the corresponding group velocity. For instance, we observe in Fig.~\ref{fig:rel_diff_E_y_linHd_halevi}(c)
that the extrema lie close to each other from about $t\approx12.03$fs (at which time
 a nice wave paket has build up) up to about $t\approx23.05$fs where the two wave 
pakets have merged. From a linear regression, we obtain the velocities $v_{\text{ue}}=0.00138c$ 
and $v_{\text{le}}=0.00139c$ (where $c$ is the vacuum speed of light) for the upper 
and lower envelope, respectively. This is a rather interesting result, since these 
velocities are considerable smaller than all velocity scales that we have discussed 
so far. In fact, the values are closest to the elastic shear velocity $\vfermi /\sqrt{5}$. 

Further, we note that within the pulse the spatial frequency experiences a chirp. 
Therefore, we provide in Fig.~\ref{fig:rel_diff_E_y_linHd_halevi}(b) the one-sided 
Fourier transform of this signal. While this provides an overview of the relevant 
contributions it lacks the desired spatial resolution. Therefore, we also provide 
in Fig.~\ref{fig:rel_diff_E_y_linHd_halevi}(b) the corresponding spectrogram. 
Although the minimal spatial resolution of the spectrogram corresponds to about 
$0.1$nm, the inverse scaling between spatial frequency and spatial resolution 
forces us to fix the spatial resolution to about $1$nm. Despite its coarseness, this
resolution allows us to elaborate on the positive spatial chirp (towards the cylinder
center) within each wave paket. Specifically, we find that the highest available 
frequencies are tied to the region, where the wave pakets are located and, in
this region, they increase towards the center. 

We conclude our survey by calculating the modal distributions for the dipolar and
quadrupolar resonance for both, the Halevi and the high-frequency Euler-Drude model 
from our time-domain simulations via an on-the-fly Fourier transform. 
In Fig.~\ref{fig:mode_pictures_driftHD_and_relative} we depict the real part of 
the electric field distributions at these resonances for the Halevi model and the
difference of these field distributions to the results of the high-frequency 
Euler-Drude model. In addition, we normalize the difference to the largest value 
obtained within the Euler-Drude model at the respective frequency in order to 
bring out better the deviations between the results.

First, we observe that these modal patterns within the Halevi model are 
compatible with the symmetry classification of the Euler-Drude model
~\cite{PhysRevB.97.075431}. 
Further, the field amplitudes in the Halevi model are enhanced. Specifically, for 
the dipole resonance we obtain an enhancement in the vicinity of the surface 
with a four-fold rotation symmetry for the $x$- and $y$-component of the electric 
field. For the quadrupolar resonance, the differences inside the scatterer are 
more inhomogeneous. Overall, the field enhancement is relatively small.

\section{\label{sec:summary_and_outlook}Conclusion and outlook}

To summarize, we have analyzed the physical properties of the Halevi model for
plasmonic materials. In particular, we have identified the Halevi  model as an
extension of the Euler-Drude model and have established its relation to the
viscoelastic model by way of spectral as well as spatio-temporal investigations 
of different aspects of plasmonic light-matter interaction. 
We have further carried out an extension of the Mie-Ruppin theory of hydrodynamic 
cylinders to the Halevi model and determined the quasi-static dispersion relation 
of surface plasmons. Thereby, we have revealed a novel damping term which exibits
formal similarities to the collision-modified Landau-damping of Halevi. In turn,
this leads to a novel physical justification of an earlier phenomenological description 
of limited-mean-free path effects of dipole surface modes and we have shown how
this novel damping term also affects higher-order surface modes.
The shear-extension within the Halevi model eventually has lead to nonlocal surface 
damping. Further, the Halevi extension to the Euler-Drude model led to an increase
in the relative difference of the width of surface dipolar and quadrupolar 
extinction peaks for cylinder radii in the range of $1$ to $10$nm, where,
generally, the effect is larger for the quadrupolar resonance.

To complement the spectral investigations, we have employed the ADE technique to 
adopt the Halevi model for time-domain simulations. The additional dispersion of 
the latter yields a modification of the diffusion current of Fick-type. We have
shown that the propagation of the induced charge, related to the current modifications, 
shares similarities with a hybrid, diffusive-wavelike paradigm as described by the 
Cattaneo equation. 

We have completed our analytical analysis by comparing the Halevi model to the
GNOR model -- the latter being another extension of the the Euler-Drude model 
that includes a diffusive contribution to the current without asymptotical 
comparison to a semiclassical model. The GNOR-model features several differences 
to the Halevi model that include a low-frequency, characteristic velocity exceeding 
the Thomas-Fermi value and the deviation from the Euler-Drude model for 
frequencies beyond the volume plasma frequency.  

Further, we have shown that the latter model is connected to a difference in the scaling 
of the diffusive length scale at plasmonic frequencies by an additional factor
$\volplasmafreq/\gamma$ relative to the Halevi model. For typical metals and 
assuming a frequency-independent GNOR-diffusion constant this additional factor
takes on values $\volplasmafreq/\gamma\sim10^2..10^3$.
As a result, while in the GNOR model the diffusive length exceeds the length scale 
of combined compression and shear, the opposite case is manifest Halevi model. 
Finally, we have shown that a direct comparison between the GNOR and the Halevi 
model at intermediate frequencies suggest that the GNOR-diffusion constant is 
complex-valued and frequency-dependent. 
 
Next, by employing the time-domain formulation of the Halevi model, we have 
numerically determined the spatio-temporal evolution of the electric fields distributions
in and around silver nano-wires. 
We have found that under pulsed excitations, the differences of the electric field 
components between the Halevi and the Euler-Drude model take on the form of wave
pakets that build up at the sub-surface region, subsequently propagates towards 
the cylinder center from where they spread across the entire cross section of the 
cylinder and eventually fade away. This behavior correlates with a concomitant 
oscillatory contribution of the diffusive current. 
Upon analyzing the resulting wave paket envelopes, we have inferred an estimate 
of the corresponding group velocity which, in fact, has turned out to be smaller 
than all natural scales of the Halevi and viscoelastic model. 
Further, we have determined a positive chirp of the wave pakets spatial frequencies
towards the center of the cylinder. 
{
In addition we have performed on-the-fly Fourier transforms of the temporal 
evolution of the field distribution and have determined the mode distributions 
of the dipolar and quadrupolar resonance. The Halevi model preserves the 
respective symmetries of the Euler-Drude model and -- at least for monomers --
the differences between both models are rather small.
}
%

Based on our results, we would like to provide a few comments regardin possible future 
routes. First of all, all our results rely on the hardwall boundary condition in the
form of the so-called slip boudary condition. On the mesoscopic scale, this constraint 
could be lifted by introducing an infinitesimal charge sheet, e.g., via a variation of 
the composite-surface model of Ref.~\onlinecite{Horovitz_2012}. 
On the one hand, this would yield an approximate treatment of spill-out effects that are
presently not contained within our model. On the other hand, it might be used to treat 
charge transfer effects at the interface to an embedding medium. This could phenomenologically 
for 'chemical interface damping', i.e., charge-transfer effects across the surface of a
plasmonic nano-particle and its surrounding host material
~\cite{PhysRevB.48.18178}. 
This could be further compared to the theory of Persson
~\cite{PERSSON1993153} 
that connects the latter to interactions with resonance states of adsorbates that form
at a metal surface. 
In addition, a detailed discussion on the difference between slip- and no-slip boundary
conditions would be highly desirable. The latter cannot be enforced ad hoc, but has to 
be physically motivated with regard to the specific surface characteristics of the 
scatterer under consideration
~\cite{PhysRevB.84.121412}. 
A potential justification could be inferred by embedding the Halevi model into the 
viscoelastic model which, in turn, would provide a clear path to the introduction of 
transverse nonlocality. Specifically, the absence of a tangential surface current due 
to e.g. surface irregularities can affect the neighborhood of the surface due to shear, 
thus providing a physical motivation of the no-slip model. 
So far, the latter has been considered only for a half-space of metal yielding a nonlocal 
correction to the well-known s-polarized Fresnel formula
~\cite{universe7040108}. 
Further, an application of the full viscoelastic model to thin semiconductor films 
has been performed in Ref.~\onlinecite{viscoel_eps_near_zero_thin_films}, 
albeit with the slip boundary condition.

Finally, for the setup of a single nano-wire considered in our work, the relative differences 
in the  widths and amplitudes of resonances between the Halevi and Euler-Drude model 
have turned out to be rather small. However, this does not always have to be the case. It is,
therefore, very interesting to seek for physical setups in which changes to the optical, 
dieletric bulk properties (as encoded in the dielectric functions) are more pronounced. 
For instance, Kreibig and Fragstein 
~\cite{limitation_elec_mean_free_path} 
have pointed out, that the extinction of colloidal metallic nanospheres is very sensitive 
to the changes of bulk quantities. Further, it might be interesting to investigate dimer 
and related structures, i.e., to consider the effect of the Halevi model on the electromagnetic 
field distribution in nano-gap systems. Recently, such systems have witnessed significant
attention due to the potential of strongly modified light-matter interaction, for instance,
with regards to strong coupling of emitters and/or enhanced nonlinear optical effects.

\begin{acknowledgements}
G.W. and K.B. acknowledge funding by the German Research Foundation (DFG) in
the framework of the Collaborative Research Center 1375 'Nonlinear Optics down 
to Atomic Scales (NOA)' (project number 398816777).
N.~A.~M. is a VILLUM Investigator supported by VILLUM FONDEN (Grant No.~16498).
The authors wish to thank Matthias Plock for fruitful discussions.

\end{acknowledgements}
\appendix* 

\section{\label{app-sec:cylindrical_surface_plasmons}Quasi-static cylindrical surface plasmons}

In this appendix, we calculate the quasi-static approximation for cylindrical surface plasmons
within the Halevi model, Eq.~(\ref{eq:halevi_slip_cyl_sp_resonance_approx}).

In Sec.~\ref{sec:Ruppin_cyl_halevi}, we have considered the scattering of a plane
wave that propagates in vacuum and is normally incident onto an infinitely extended 
circular cylinder for the polarization perpendicular to the cylinder axis.
Via Ruppin's extension of the correspinding Mie solution (expansion into spherical
harmonics), we have obtained the expansion coefficients for of the scattered field 
in Eq.~(\ref{eq:Ruppin-scatt-coeff}). The surface resonances can be determined from
the poles of expansion coefficients according to  
\begin{align}
   0 &= \left[c_n  + D_n(k_{\rm T}a)\right]H_n(k_0 a ) - \sqrt{\epsilon_{\rm T}(\omega)}H'_n(k_0a).
\label{eq:Ruppin_scatt_coeff_pole}
\end{align}
As an illustration, we consider silver with material constants given by 
Tab.~\ref{tab:metal_material_params_and_damping}. From the definitions of the 
three wave numbers $k_0, k_{\rm L}, k_{\rm T}$, we deduce for cylinders with
radii $a=1..10$nm that
\begin{align}
|k_0a|,\, |k_{\rm T}a| \lesssim 0.1 \,\,\, \text{and} \,\,\, 10 \lesssim |k_{\rm L}a|.
\end{align}
Accordingly, we may utilize of Eq.~(\ref{eq:Ruppin_scatt_coeff_pole}) the Bessel functions' 
asymptotic representations. We then find the implicit dispersion relation
\begin{align}
   \epsilon_{\rm T}(\omega) &= 
	                         -   \frac{1 - in / k_{\rm L}a }{1 + in / k_{\rm L}a } 
					                  \approx 
														   -1 + \frac{2in}{k_{\rm L}a}, 
\label{eq:implicit_disp_rel_Halevi_cyl_surf_modes}
\end{align} 
where, in the last step, we have further assumed, that $n \ll |k_{\rm L} a|$. 
Using Eq.~(\ref{eq:Drude_diel_func}), we find that the local solution to first order 
in $\gamma/\omega_{\rm p}$ assumes the form
\begin{align}
   \omega_{\rm loc} & = \omega_{\rm sp} - \frac{i\gamma}{2} 
	                      + \order{\frac{\gamma^2}{\surfplasmafreq^2}}.
\label{eq:loc_disp_rel_of_quasi_stat_cylinder}
\end{align}
Here, we have selected the solution with a positive real part. Note, that this is also 
the local dispersion relation for surface plasmons at a planar interface between 
vacuum and a half-space filled with metal. 
Consequently, for normally incident and quasistatic electic fields polarized perpendicular 
to the cylinder axis, the geometry does not alter the local surface plasmons. This is 
a consequence of the quasi-static approximation. Otherwise, even the local solution 
would depend on the pole order $n$ and radius $a$.

Next, we seek a solution to Eq.~(\ref{eq:implicit_disp_rel_Halevi_cyl_surf_modes}), 
that includes nonlocal correction to Eq.~(\ref{eq:loc_disp_rel_of_quasi_stat_cylinder}) 
to leading order. Therefore. we use the Ansatz
\begin{align}
   \omega_n &= 
	            \surfplasmafreq \left(1 - \frac{i\gamma}{2\surfplasmafreq} + f_n\right) \,\,\, 
							\text{with} \,\,\, |f_n| \ll 1.
\end{align}
Considering the Taylor expansion in $f_n$ of Eq.~(\ref{eq:Drude_diel_func}) around 
$\omega=\omega_n$, we find
\begin{align}
   \epsilon_{\rm T}(\omega_n) = \epsilon_{\rm T}(\tildesurfplasmafreq) 
	                              + \epsilon'_{\rm T}(\tildesurfplasmafreq) \surfplasmafreq f_n + \order{f^2_n},
\label{eq:epsilon_T_Taylor_in_f_n}
\end{align}
where we have introduced $\tildesurfplasmafreq=\surfplasmafreq - i \gamma / 2$.
To make further progress, we require the following Taylor expansions in $\gamma/\surfplasmafreq$
\begin{align}
   \epsilon_{\rm T}(\tildesurfplasmafreq)  & = -1 + \order{\frac{\gamma^2}{\surfplasmafreq^2}} \,\,\, 
	 \nonumber \\ 
   \epsilon'_{\rm T}(\tildesurfplasmafreq) & = \frac{4}{\surfplasmafreq} + \order{\frac{\gamma^2}{\surfplasmafreq^2}},
\label{eq:epsilon_T_and_deriv_Taylor_in_gamma_omega_sp}
\end{align}
which yields 
\begin{align}
   \epsilon_{\rm T}(\omega_n) \approx -1  + 4 f_n.
\label{eq:epsilon_T_doubel_Taylor}
\end{align}
Equating this result with the r.h.s. of Eq.~(\ref{eq:implicit_disp_rel_Halevi_cyl_surf_modes}), 
we find for the nonlocal correction the implicit relation
\begin{align}
   f_n &= \frac{in}{2a} \frac{1}{k_{\rm L}(\omega_n)}.
\label{eq:implicit_relation_f_n}
\end{align}

Keeping Eq.~(\ref{eq:longitudinal_wavenumber_square_Halevi}) in mind, we have to expand 
Eq.~(\ref{eq:halevi_beta_square}) in a manner analoguous to 
Eqs.~(\ref{eq:epsilon_T_Taylor_in_f_n}, \ref{eq:epsilon_T_and_deriv_Taylor_in_gamma_omega_sp}) 
and eventually arrive at
\begin{align}
   \beta^2_{\rm H}(\omega_n) \approx \left[1 - \frac{4i}{9}\frac{\gamma}{\surfplasmafreq}\right] \beta^2_{\rm HF}
\label{eq:halevi_beta_sq_double_Taylor}
\end{align}
Here, we have dropped the first-order term of the Taylor expansion because this is proportional 
to $(\gamma/\omega_{\rm sp})f_n$.

Upon inserting Eqs.~(\ref{eq:epsilon_T_doubel_Taylor}, \ref{eq:halevi_beta_sq_double_Taylor}) 
into Eq.~(\ref{eq:longitudinal_wavenumber_square_Halevi}) we obtain 
\begin{align}
   \frac{1}{k_{\rm L}(\omega_n)} \approx 
	                               -\frac{i\beta_{\rm HF}}{\surfplasmafreq}
																 \left[1 - \frac{2i}{9}\frac{\gamma}{\surfplasmafreq} + f_n\right]
\end{align}
which, when inserted into Eq.~(\ref{eq:implicit_relation_f_n}), gives
\begin{align}
   f_n = \frac{n\beta_{\rm HF}}{2\surfplasmafreq a}
	       \left(1 - \frac{2i}{9}\frac{\gamma}{\surfplasmafreq}\right) 
				 \left[1 - \frac{n\beta_{\rm HF}}{2\surfplasmafreq a}  \right]^{-1}
\end{align}
Indeed, since $n \ll \surfplasmafreq a/\beta_{\rm HF}$ and $\gamma\ll \surfplasmafreq$, we find 
$|f_n|\ll1$. To first order in $n\beta_{\rm HF}/\surfplasmafreq a$, the second term inside the 
square brackets can be neglected and this gives Eq.~(\ref{eq:halevi_slip_cyl_sp_resonance_approx}). 


\end{document}